\documentclass[aps,preprint,twocolumn,notitlepage,superscriptaddress,10pt,floatfix,a4paper,nofootinbib]{revtex4-1}
\pdfoutput=1 
\usepackage{xcolor}
\usepackage{enumitem}
\usepackage{graphicx}
\usepackage{stmaryrd}
\usepackage{amssymb,amsfonts}
\usepackage{amsmath}
\usepackage{mathrsfs}
\usepackage{esint}
\usepackage{bbm}
\usepackage{enumerate}
\usepackage{verbatim}
\usepackage{subfigure}
\usepackage{hyperref}
\newcommand{\bra}[1]{\langle{#1}|}
\newcommand{\ket}[1]{|{#1}\rangle}
\newcommand{\braket}[2]{\langle{#1}|{#2}\rangle}
\renewcommand{\d}[1]{\ensuremath{\operatorname{d}\!{#1}}}
\renewcommand{\vec}[1]{\boldsymbol{#1}}

\begin{document}
\title{Getting coherent state superpositions to stay put in phase space:\texorpdfstring{\\ 
$Q$}{Q} functions and one dimensional integral representations of generator eigenstates}
\author{Mayukh Nilay Khan}
\email{hereismayukh@gmail.com}
\date{\today}
\begin{abstract}
We study quantum mechanics in the phase space associated with the coherent state (CS) manifold of Lie groups.
Eigenstates of generators of the group are constructed as one dimensional integral superpositions of
CS along their orbits. 
We distinguish certain privileged orbits where the superposition is in phase.
Interestingly, for closed in phase orbits, the geometric phase must be quantized to $2\pi\mathbb{Z}$, else the superposition vanishes. 
This corresponds to exact Bohr-Sommerfeld quantization.
The maximum of the Husimi-Kano $Q$ quasiprobability distribution is used to diagnose where in phase space the 
eigenstates of the generators lie. The $Q$ function of a generator eigenstate is constant along each orbit.
We conjecture that the maximum of the $Q$ function corresponds to these privileged in phase orbits.
We provide some intuition for this proposition using interference in phase space,
and then demonstrate it for
canonical CS ($H_4$ oscillator group), spin CS ($SU(2)$) and $SU(1,1)$ CS, relevant to squeezing. 
\end{abstract}
\maketitle
\section{Introduction}
The study of coherent states has become ubiquitous in different branches of physics. First introduced by Schr\"{o}dinger
\cite{schrodinger1926stetige}
while investigating quantum mechanical wave packets whose evolution most closely approximate
the classical dynamics of the simple harmonic oscillator (SHO), coherent states (CS) have now been generalized 
so that their applicability extends to arbitrary Lie groups.\cite{perelomov1972coherent}

In this article, we start with the construction of eigenstates of Lie group generators as superpositions along their corresponding 
orbits in the CS manifold. We see that geometric phase \cite{Berry,Pancharatnam,kinematicformulation1,kinematicformulation2}
ideas appear naturally in the process (especially for closed orbits). We obtain 
integral quantization of geometric phase along in phase closed orbits.
After this we consider some interesting facets of this construction as outlined below. 

The manifold of generalized CS (with appropriate choice of fiducial state) can be treated as a phase space 
with appropriate symplectic strucure in a number of typical situations. We start with the question of where a state $\ket{\psi}$ 
is located in phase space. If a point in phase space is parametrized by $\vec\eta$,
and hence the associated CS is $\ket{\vec\eta}$, a simple diagnostic would be the regions where $\vert\braket{\vec\eta}{\psi}\vert^2$ is maximized. Indeed,
this is the Husimi-Kano $Q$ quasiprobability distribution \cite{HusimiQ} in phase 
space.
\begin{align}
Q_{\ket{\psi}}(\vec\eta)=\vert\braket{\vec\eta}{\psi}\vert^2
\label{eqn:qfunc}
\end{align}
We illustrate our main ideas in the context of the familiar SHO CS. 
(The notation used in this section is standard, for elaboration refer to main text, Section~\ref{subsec:SHOcoherentstates}.)

Let us start with the CS $\ket{q,p}$.
The $Q$ function 
$Q_{\ket{q,p}}(q',p')=e^{-\frac{1}{2}\left(q-q'\right)^2-\frac{1}{2}\left(p-p'\right)^2}$. This matches our intuitive notion that the CS
is localized at $(q,p)$.
The uncertainty principle precludes arbitrarily narrow distributions in phase space. 
However, in a quantifiable sense CS are the most localized states
\cite{wehrl1979relation,gnutzmann2001renyi}
in phase space. Wehrl conjectured \cite{wehrl1979relation} that among the set of density matrices $\hat\rho$ in 
a Cartesian quantum mechanical system, projectors onto coherent states  minimize the Wehrl entropy $S_W$. 
\[S_W(\hat{\rho})=-\iint \frac{dqdp}{2\pi} Q(q,p)\log Q(q,p),\]
where $Q(q,p)=\bra{q,p}\hat{\rho}\ket{q,p}$ is the $Q$ function of $\hat{\rho}$.
This was proved soon after by Lieb. \cite{lieb1978proof}
Actually, they are the unique states that minimize the Wehrl entropy \cite{CARLEN1991231}, and this result extends to
Bloch CS too.
\cite{lieb2014proof}

Thus, while one expects the CS $\ket{\vec\eta}$ to be localized at the corresponding point in phase space, what about other states?
In this article, we restrict ourselves to eigenstates of the generators of the group. We conjecture that the state is localized/$Q$ function maximized
along special ``in phase" orbits. We also construct the eigenstate as a superposition over such orbits. This superposition stays where it is put.

In particular, consider the Fock state $\ket{m}$ which is an eigenstate of the number operator $\hat{n}$, a generator of the $H_4$ oscillator
group. Now, it is well known that a Fock state can be expanded  as a superposition of CS over arbitrary circles in phase space
(orbits generated by $\hat{n}$).
\begin{align}
&\ket{m}=\frac{\sqrt{m!}}{2\pi}\left(\frac{\sqrt{2}}{r}\right)^m e^{\frac{r^2}{4}}\int_0^{2\pi}\d\theta\,e^{-im\theta}\,
\ket{r\cos\theta,r\sin\theta}.
\label{eqn:introFockQstate}
\end{align}
We note the following points about the expression above.
\begin{itemize}[leftmargin=*]
\item $\ket{m}$ can be expressed as a superposition over an arbitrary circle. (any orbit of $\hat{n}$)
\item The normalizing coefficient outside the integral is {\emph{minimized}} at $r=\sqrt{2m}$. 
We refer to this as the superposition being ``efficient". The further away we choose the orbit from the ``natural" location of $\ket{m}$
at $r=\sqrt{2m}$, the larger the
normalizing coefficient.
\item Over the ``privileged" orbit at $r=\sqrt{2m}$ the geometric phase is quantized to $2\pi m$
(we will see that this is characteristic of superpositions over closed orbits), and the superposition is ``horizontal/in phase".
These statements will be explained in detail in the main sections.	
\item The $Q$ function of $\ket{m}$ (Eqn.~\eqref{eqn:QdistributionH4}) is constant on any circle. However, its value is 
{\emph{maximized}} on the circle of radius $\sqrt{2m}$.

Thus, although we can express $\ket{m}$ as a superposition over any circle, its natural position as diagnosed by the $Q$ function
is at $r=\sqrt{2m}$. A superposition over $r=\sqrt{2m}$ stays in place. 
\item Finally, at $r=\sqrt{2m}$ we have exact Bohr-Sommerfeld condition of old
quantum theory $\oint p\,dq=2\pi m$. 
\end{itemize}
It turns out that these features are true for generalized systems of coherent states (GCS) with some assumptions, as we will see in the succeeding
sections.

We note that in a talk given in 2001~\cite{simontalk}, R. Simon had presented
generator eigenstates as 
group averages over their in phase orbits.
Further, he also emphasized the importance of geometric phase in analyzing interference patterns
in the $Q$ distribution due to superpositions of coherent states.
We present some of these ideas and develop them further in Secs.~\ref{subsec:constructiongeneral},\,\ref{sec:intuition}.

In Ref.~\cite{khan2018geometric}, these ideas 
have been discussed at length for SHO CS in the context of understanding 
interference in phase space \cite{Wheeler_1985,Schleich_1987} from a geometric phase perspective without
using any semiclassical arguments. 
Here, we follow the same logical steps as Ref.~\cite{khan2018geometric}, stating these results 
so that they are valid for general CS systems, but concentrating on the conjecture about $Q$ function maximization for in phase orbits.
\section{General Formalism}\label{sec:general formalism}
\subsection{Generalized coherent states}
We consider generalized coherent states as introduced in Refs. 
\cite{perelomov1972coherent,CoherentPerelomov}. Let $G$ be a 
Lie group and $T$ its irreducible unitary representation acting on
the Hilbert space $\mathcal{H}$. We choose a fixed fiducial state $\ket{0}$ in $\mathcal{H}$. 
Then, the coherent state (CS) system $\{G,T,\ket{0}\}$
corresponds to the set of states
$\{\ket{\alpha_g}\},\,g\in G$, where $\ket{\alpha_g}={T}(g)\ket{0}$.
However, note that $\ket{\alpha_{g_1}}$ and $\ket{\alpha_{g_2}}$ differ only by a $U(1)$ phase if
${T}(g_2^{-1}g_1)\ket{0}=e^{i\phi}\ket{0}$, and hence represent the same physical state.
To eliminate this redundancy, one defines the subgroup $H$,
${T}(h\in H)\ket{0}=e^{i\phi(h)}\ket{0}$.
The maximal subgroup $H$ is called the isotropy subgroup of $\ket{0}$.
The CS manifold is indexed by elements $x$ of the left coset space $\Omega=G/H$. 
Thus, ${T}(g)\ket{0}=e^{i\phi(g)}\ket{x(g)}$, where $x\in G/H$ is the 
equivalence class to which $g$ belongs. $\Omega$ is the ``phase space" of our quantum mechanical system.

While any state in $\mathcal{H}$ can be chosen as the fiducial state, if $\ket{0}$ is chosen to have the largest 
isotropy subalgebra, the CS are closest to classical states and have minimum uncertainty
\cite{gilmore1974properties,delbourgo1977maximum,CoherentPerelomov,GilmoreCoherentReview}.
With such choice of $\ket{0}$, $\Omega$ is endowed with a symplectic structure, 
in fact they are complex homogenous manifolds with a K\"{a}hler structure in a number of typical cases.\cite{Onofri_1975}
( Compact/solvable Lie groups and also some non-compact cases with discrete representations. For details refer to
\cite{Onofri_1975,CoherentPerelomov}.)

Henceforth,
we denote the operator $\mathscr{O}$ in the representation $T$ acting on $\mathcal{H}$ i.e. $T(\mathscr{O})$ as $\mathscr{\hat{O}}$.

\subsection{Geometric phase preliminaries}
The study of geometric phases has played an important role in different branches of physics ever since it was 
introduced by Berry in the context of adiabatic cyclic quantum evolution \cite{Berry}, where it is
called the Berry phase. 

We use the kinetic formulation \cite{kinematicformulation1,kinematicformulation2} where the restrictions of adiabaticity and cyclic evolution
can be revoked \cite{adiabaticity,cyclicevolution}.
Given a once differentiable parametrized curve $\mathscr{C}(\chi)$ in the space of normalized states in $\mathcal{H}$, 
we define the relative or Pancharatnam phase $\phi_P$, 
the dynamical phase $\phi_d$ and the geometric phase $\phi_{\text{geom}}$
as we move from $\chi_1$ to $\chi_2$ as
\begin{align}
	&\mathscr{C}=\{\ket{\psi(\chi)}\,:\,\, \braket{\psi(\chi)}{\psi(\chi)}=1;\,\,\chi_1\leq \chi\leq \chi_2\,\};\nonumber\\
&\phi_P(\mathscr{C})=\arg\,\braket{\psi(\chi_1)}{\psi(\chi_2)};\label{eqn:panchadefn}\\
&\phi_{\text{dyn}}(\mathscr{C})=-i\int_{\chi_1}^{\chi_2}\d \chi\, \braket{\psi(\chi)}{\frac{\d\psi(\chi)}{\d \chi}};\label{eqn:dynphasedefn}\\
&\pi(\mathscr{C})=C=\{\,\hat{\rho}_{\ket{\chi}}=\ket{\psi(\chi)}\bra{\psi(\chi)}\,;\;\,\,\chi_1\leq \chi\leq \chi_2\,\};
\label{eqn:rayspaceprojectiongeneral}\\
&\phi_{\text{geom}}(C)=\phi_P(\mathscr{C})-\phi_{\text{dyn}}(\mathscr{C}).
\label{eqn:geomphasedefn}
\end{align}
It is easy to check that $\phi_{\text{geom}}$ is invariant under local $U(1)$ transformations, and 
hence solely depends on the 
projection $\pi(\mathscr{C})=C$ in ray space $\mathcal{R}$.
\begin{align*}
	\mathcal{R}=\{\hat\rho_{\ket{\psi}}=\ket{\psi}\bra{\psi}:\,\,\braket{\psi}{\psi}=1\}. 
\end{align*}
$\phi_{\text{geom}}$ being a property of projective ray space $\mathcal{R}$, to calculate it
we can transform to the ``locally" in phase/horizontal curve $\mathscr{C}_{\text{hor}}$, where 
neighbouring points are in phase with each other.
\begin{align}
	&\mathscr{C}_{\text{hor}}=\{\ket{\psi'(\chi)}: \ket{\psi'(\chi)}=\ket{\psi(\chi)}\,e^{i\phi(\chi)},\,\chi_1\leq \chi\leq \chi_2\},\nonumber\\
	&\phi(\chi)=i\int_{\chi_1}^{\chi}\d \chi'\bra{\psi(\chi')}\frac{d}{d\chi'}\ket{\psi(\chi')}.\label{eqn:horizontalcurvegeometricphase}\\
	&\phi_{\text{dyn}}(\mathscr{C}_{\text{hor}})=0,~ \phi_{\text{geom}}(C)=\arg\braket{\psi'(\chi_1)}{\psi'(\chi_2)}.  \nonumber
\end{align}
We briefly discuss two related ideas which actually precede Ref.~\cite{Berry} and will be of some use to us.

$\ket{\psi_1}$ and $\ket{\psi_2}$ are in phase if
$\braket{\psi_1}{\psi_2}\in\mathbb{R}^+$. 
In general, the relative phase between $\ket{\psi_1}$ and $\ket{\psi_2}$ is
\begin{align}
	\phi_P(\ket{\psi_1},\ket{\psi_2})=\arg\braket{\psi_1}{\psi_2}.
\label{eqn:pancharelphase}
\end{align}
While studying polarization states of unidirectional light,
Pancharatnam \cite{Pancharatnam,BerryPancharatnam} realized that {{being in phase is not a transitive relation}},
and he gave a measure of the 
non-transitivity.
Distinct polarizations are specified by $\hat\rho_a=\ket{\psi_a}\bra{\psi_a}$, (where $\ket{\psi_a}$ is a 2 component
complex electric field vector) and correspond to points on the Poincar\'e sphere $S^2$.
If $\ket{\psi_a}$ and $\ket{\psi_b}$ are in phase, and so is $\ket{\psi_b}$ and $\ket{\psi_c}$, the phase difference between 
$\ket{\psi_c}$ and $\ket{\psi_a}$ has a simple geometrical interpretation.
$\arg\braket{\psi_a}{\psi_c}=-\frac{1}{2}\Omega_{\hat{\rho}_{a},\hat{\rho}_{b},\hat{\rho}_{c}}$, where
$\Omega_{\hat{\rho}_{a},\hat{\rho}_{b},\hat{\rho}_{c}}$ is the solid angle subtended by the spherical triangle 
$(\hat{\rho}_{a},\hat{\rho}_{b},\hat{\rho}_{c})$ on $S^2$.

The relative phase between $\ket{\psi_a}$ and $\ket{\psi_c}$ can also be expressed as
\begin{align}
&\arg\braket{\psi_a}{\psi_c}=-\arg\braket{\psi_a}{\psi_b}\braket{\psi_b}{\psi_c}\braket{\psi_c}{\psi_a}\nonumber\\
&=-\arg\,\mbox{Tr}\left(\hat{\rho}_{a}\hat{\rho}_{b}\hat{\rho}_{c}\right)
 =-\text{arg}\Delta_3\,(\hat{\rho}_{a},\hat{\rho}_{b},\hat{\rho}_{c}),\label{eqn:Bargmanninvariant}
\end{align}
where $\Delta_3\,(\hat{\rho}_{a},\hat{\rho}_{b},\hat{\rho}_{c})=\mbox{Tr}(\hat{\rho}_{a}\hat{\rho}_{b}\hat{\rho}_{c})$ 
is the three vertex Bargmann invariant introduced 
\cite{Bargmanninvariant} in the context of proving Wigner's theorem on symmetries in quantum mechanics, and like $\phi_{\text{geom}}$
is obviously a property of ray space $\mathcal{R}$.
As we will see, $\Delta_3$ is intimately related to the geometric phase.

We now state a few results for Hilbert space curves over the CS manifold $\Omega$. Consider the closed curve in $\Omega$
\begin{align}
	\mathfrak{C}=\{{\alpha(\chi)}:~{\alpha(\chi_0)}={\alpha(0)},\,0\leq \chi<\chi_0\},
\label{eqn:closedcoherentmanifoldcurve}
\end{align}
and the corresponding curve in $\mathcal{H}$
\begin{align}
	\mathscr{C}=\{\ket{\alpha(\chi)}:~\ket{\alpha(\chi_0)}=\ket{\alpha(0)},\,0\leq \chi<\chi_0\}.
\label{eqn:closedcoherentcurve}
\end{align}
When $\Omega$ is K\"{a}hler, the geometric phase for cyclic quantum motion along $\mathscr{C}$ is equal to the negative 
of the symplectic area enclosed by $\mathfrak{C}$
in $\Omega$. \cite{boya2001berry} 
We see this in the examples in Section~\ref{sec:examples}.

The following generalization of Pancharatnam's result on non-transitivity stated in terms of the Bargmann invariant holds true 
for a certain sub-class of K\"{a}hler manifolds, the so called ``Hermitian symmetric" manifolds. Actually, a number of familiar manifolds
are Hermitian symmetric (the complex plane, Riemann sphere $\mathbb{C}P^1$ and Poincar\'{e} disk are examples we will encounter in this article).

Consider the geodesic triangle ($g_1,g_2,g_3$) in Hermitian symmetric $\Omega$. 
The geometric phase for the circuit
$\ket{g_1}\shortrightarrow\ket{g_2}\shortrightarrow\ket{g_3}\shortrightarrow\ket{g_1}$ along the geodesic triangle is
\begin{align}
&\phi_{\text{geom}}(\ket{g_1}\shortrightarrow\ket{g_2}\shortrightarrow\ket{g_3}\shortrightarrow\ket{g_1})=-\mathscr{A}(g_1,g_2,g_3)\label{eqn:symplecticgeodesictriangle}\\
&=-\arg\Delta_3\,(\hat\rho_{g_1},\hat\rho_{g_2},\hat\rho_{g_3}),\label{eqn:BargmannInvariantGCS}
\end{align}
where $\mathscr{A}(g_1,g_2,g_3)$ is the symplectic area of the geodesic triangle
(Refs.~\cite{boya2001berry,domictoledo1987,clerc2003corrigendum,berceanu1999coherent,berceanu2004geometrical,bech2017canonical}).
In Eqn.~\eqref{eqn:BargmannInvariantGCS}, neighbouring points on the sequence are assumed to be non-orthogonal,
$\braket{g_{i+1}}{g_i}\neq 0,\,\braket{g_1}{g_3}\neq 0$.

For a triad of points on a geodesic, the area enclosed vanishes. Thus, on geodesics being in phase in an equivalence relation,
and is a global statement.
While considering geodesics in the projective ray space 
$\mathcal{R}$, similar relations and more generalized versions \cite{kinematicformulation1,Rabei_1999,mukunda2003}
have been explored.

\subsection{Construction of eigenstates of generators}\label{subsec:constructiongeneral}
Let $\mathcal{T}$ be a generator of the group $G$, i.e. 
$\mathcal{T}\in\mathfrak{g}$, the Lie algebra.
The action of the one parameter
subgroup $e^{-i\chi\mathcal{T}}$, generated by $\mathcal{T}$, 
foliates $\Omega$ into orbits of $\mathcal{T}$. 
$\mathcal{O}_{\mathcal{ T},\,\alpha_0}$ is the group orbit of $\mathcal{ T}$
in $\Omega$ passing through $\alpha_0$. 
\begin{align}
	\mathcal{O}_{\mathcal{ T},\,\alpha_0}=\{ e^{-i\chi\mathcal{ T}}\alpha_0\,\vert\,\alpha_0\in\Omega,\,\mathcal{ T}\in\mathfrak{g}\}.
\label{eqn:classicalorbit}
\end{align}
Above, the limits of $\chi$ are not explicit. For open orbits $-\infty<\chi<\infty$, 
whereas for closed orbits the bounds are finite, $0\leq\chi<\chi_0$.

We now construct eigenstates of the Hilbert space operator $T(\mathcal{T})\equiv \mathcal{\hat T}$ as a one dimensional integral 
superposition of CS over (almost) any given orbit
generated by $\mathcal{T}$ in $\Omega$. When we fail to do so, $\ket{\alpha_0}$ is an eigenstate of $\mathcal{\hat T}$
to begin with and $\mathcal{O}_{\mathcal{ T},\,\alpha_0}$ is just the point $\alpha_0$.
(For example consider the orbit of the number operator $\hat{n}$ acting on the ground state $\ket{0}$ of the SHO CS).

This should come as no suprise, as CS are in general overcomplete.
For oscillator CS,
any state in $\mathcal{H}$ can be represented as an expansion over 
``characteristic sets" e.g. smooth curves of finite or infinite length, or an
infinite sequence of points with a finite limit point \cite{bargmann1961hilbert,bargmann1971completeness}.

Corresponding to the orbit $\mathcal{O}_{\mathcal{ T},\,\alpha_0}$ in the group manifold, we consider the family of Hilbert space curves 
$\mathscr{C}_{\mathcal{\hat T},\,\ket{\alpha_0}}(t_0)$ 
\begin{align}
&\mathscr{C}_{\mathcal{\hat T},\,\ket{\alpha_0}}(t_0)=\{\,\ket{\psi_\chi(t_0)}=e^{i\chi(t_0-\hat{\mathcal{ T})}}\ket{\alpha_0}\,\},
\label{eqn:generalLiftorbit}
\end{align}
with limits on $\chi$ left implicit as before.

In general, the state $e^{-i\chi\mathcal{\hat{T}}}\ket{\alpha_0}$ differs from the CS 
$\ket{e^{-i\chi\mathcal{T}}\alpha_0}$ (or $\ket{x(e^{-i\chi\mathcal{T}}\alpha_0)}$ )
corresponding to $e^{-i\chi\mathcal{T}}\alpha_0\in\mathcal{O}_{\mathcal{ T},\,\alpha_0}$,
by a $U(1)$ phase. 

Indeed, $\pi(\mathscr{C}_{\mathcal{\hat T},\,\ket{\alpha_0}}(t_0))$ is independent of $t_0$.
\[\pi(\mathscr{C}_{\mathcal{\hat T},\,\ket{\alpha_0}}(t_0))=\{e^{-i\chi\mathcal{\hat{T}}}\hat{\rho}_{\ket{\alpha_0}}e^{i\chi\mathcal{\hat{T}}}
=\ket{e^{-i\chi\mathcal{ T}}\alpha_0}\bra{e^{-i\chi\mathcal{ T}}\alpha_0}\,\}.\]
Thus,
$\pi(\mathscr{C}_{\mathcal{\hat T},\,\ket{\alpha_0}}(t_0))$ is essentially specified by the group orbit 
$\mathcal{O}_{\mathcal{ T},\,\alpha_0}$.

Further, since geometric phase is a property of the ray space projection 
$\pi(\mathscr{C}_{\mathcal{\hat T},\,\ket{\alpha_0}}(t_0))$, $\phi_\text{geom}(\mathscr{C}_{\mathcal{\hat T},\,\ket{\alpha_0}}(t_0))$
is independent of $t_0$ and is determined by $\mathcal{O}_{\mathcal{ T},\,\alpha_0}$.

To determine whether $\mathcal{O}_{\mathcal{ T},\,\alpha_0}$ is open or closed 
in $\Omega$, we  need only look at
$\pi(\mathscr{C}_{\mathcal{\hat T},\,\ket{\alpha_0}}(t_0))$.  
$\mathcal{O}_{\mathcal{ T},\,\alpha_0}$
is closed iff 
\[\exists \chi_0:\ket{\psi_{\chi_0}(t_0)}\bra{\psi_{\chi_0}(t_0)}=\ket{\alpha_0}\bra{\alpha_0}.\] 
The smallest 
positive value $\chi_0$ is called the period of the generator. Henceforth, we will reserve the symbol $\chi_0$ for the period.

To construct eigenstates for open orbits, 
we simply average over the curve $\mathscr{C}_{\mathcal{\hat T},\,\ket{\alpha_0}}(t_0)$:
\begin{align}\label{eqn:generalopenorbitaverage}
&\ket{\bar{\psi}_{\mathcal{\hat T},\,t_0}}
=N_{\alpha_0,\,t_0}\int_{-\infty}^{\infty}\d\chi\,\ket{\psi_\chi(t_0)}.
\end{align}
$N_{\alpha_0,\,t_0}$ is the normalization factor. 
One can easily verify
\begin{align}\label{eqn:Eigenvaleqn}
e^{-i\epsilon\mathcal{\hat T}}\ket{\bar{\psi}_{\mathcal{\hat T},\,t_0}}&=e^{-it_0\epsilon}\ket{\bar{\psi}_{\mathcal{\hat T},\,t_0}}\\
\implies\mathcal{\hat T}\ket{\bar{\psi}_{\mathcal{\hat T},\,t_0}}&=t_0\,\ket{\bar{\psi}_{\mathcal{\hat T},\,t_0}},\nonumber 
\end{align}
thus concluding that $\ket{\bar{\psi}_{\mathcal{\hat T},\,t_0}}$ is an eigenstate of $\mathcal{\hat T}$ with eigenvalue $t_0$.

The Pancharatnam phase between two infinitesimally separated points along $\mathscr{C}_{\mathcal{\hat T},\,\ket{\alpha_0}}(t_0)$ is
\begin{align}
\phi_P(\ket{\psi_{\chi}(t_0)},\ket{\psi_{\chi+\delta\chi}(t_0)})&=\nonumber\\
t_0\delta\chi+\arg\bra{\alpha_0}e^{-i\delta\chi\hat{\mathcal{T}}}\ket{\alpha_0}
&=(t_0-\langle\mathcal{\hat T}\rangle)\,\delta\chi.
\label{eqn:PanchaPhaseDiff}
\end{align}
Here, $\langle\mathcal{\hat T}\rangle$ is the same for all states along $\mathscr{C}_{\mathcal{\hat T},\,\ket{\alpha_0}}(t_0)$,
and is determined by $\ket{\alpha_0}$. 
\[\bra{\alpha_0}\mathcal{\hat T}\ket{\alpha_0}=\bra{\psi_\chi (t_0)}\mathcal{\hat T}\ket{\psi_\chi (t_0)}=\langle\mathcal{\hat T}\rangle
.\]
Thus, from Eqn.~\eqref{eqn:PanchaPhaseDiff}, when 
\begin{align}
t_0=\langle\mathcal{\hat T}\rangle,
\label{eqn:localinphasesuperpositiongeneral}
\end{align}
neighbouring states are in phase with each other, and 
we have a local in phase superposition $\mathscr{C}_{\mathcal{\hat T},\,\ket{\alpha_0}}(\langle\hat{\mathcal{T}}\rangle)$.

The dynamical phase as we move along $\mathscr{C}_{\mathcal{\hat T},\,\ket{\alpha_0}}(t_0)$ between 
$\chi:\chi_1\rightarrow\,\chi_1+\eta$ is (using 
Eqn.~\eqref{eqn:dynphasedefn}) 
\begin{align}
	\phi_{\text{dyn}}(\chi_1\rightarrow\chi_1+\eta)&=-i\int_{\chi_1}^{\chi_1+\eta}\d\chi\,\bra{\psi_\chi(t_0)}\frac{\text{d}}{\text{d}\chi}\ket{\psi_\chi(t_0)}\nonumber\\ 
		 &=(t_0-\langle\mathcal{\hat T}\rangle)\eta.
\label{eqn:dynphasealongorbit}
\end{align}
Thus the Pancharatnam, dynamical and hence the geometric phases are 
translationally invariant, they do not depend on the starting point but rather on the spacing between the states. Therefore, 
we use the notation $\phi_{\text{dyn}}(\eta)$ (and similarly for $\phi_{\text{geom}}$ and $\phi_P$)
instead of $\phi_{\text{dyn}}(\chi_1\rightarrow\chi_1+\eta)$. 

For closed orbits, the integration is only over the period $\chi_0$. Thus,
\begin{align}
	\ket{\bar{\psi}_{\mathcal{\hat T},\,t_0}}&=N_{\alpha_0,\,t_0}\int_0^{\chi_0}\d\chi\, \ket{\psi_{\chi}(t_0)}.
\label{eqn:averageclosedorbit}
\end{align}
As we will see now, other constraints need to be satisfied. 

(To be precise, at this point that we have tacitly assumed that the group average formula
Eqn.~\eqref{eqn:generalopenorbitaverage}, valid for the open orbit case is valid for the closed orbit too.)
\begin{align*}
	&	\text{Substituting Eqn.~\eqref{eqn:averageclosedorbit} in }~~e^{-i\mathcal{\hat T}\epsilon} \ket{\bar{\psi}_{\mathcal{\hat T},\,t_0}}=e^{-it_0\epsilon} \ket{\bar{\psi}_{\mathcal{\hat T},\,t_0}},
\end{align*}
we get
\begin{align}
&\int_{0}^{\chi_0}\d\chi\,\ket{\psi_\chi(t_0)}
=\int_{\epsilon}^{\chi_0+\epsilon}\d\chi\,\ket{\psi_\chi(t_0)}.
\label{eqn:independentaverageclosedorbit}
\end{align}
Essentially, the average $\ket{\bar{\psi}_{\mathcal{\hat T},\,t_0}}$ is independent of the starting point $\epsilon$. This is only possible when 
the integrand is single-valued.  Using Eqns.~\eqref{eqn:geomphasedefn} and \eqref{eqn:dynphasealongorbit}, we get
\begin{align}
&\phi_P(\ket{\psi_0(t_0)},\ket{\psi_{\chi_0}(t_0)})=\phi_{\text{geom}}(\chi_0)+\phi_{\text{dyn}}(\chi_0)=2\pi m, \nonumber\\ 
&\implies\phi_{\text{geom}}(\chi_0)+\chi_0\,(t_0-\langle\mathcal{\hat T}\rangle)=2\pi m;\,m\in\mathbb{Z}. 
\label{eqn:quantiationclosedorbits1}
\end{align}
Alternatively, we can derive Eqn.~\eqref{eqn:quantiationclosedorbits1} more concretely by setting $\epsilon=\chi_0$ in 
Eqn.~\eqref{eqn:independentaverageclosedorbit}
\begin{align}
	\ket{\bar{\psi}_{\mathcal{\hat T},\,t_0}}&=N\int_{\chi_0}^{2\chi_0}\d\chi\, \ket{\psi_{\chi}(t_0)}
	=N\int_0^{\chi_0}\,\d\chi\, \ket{\psi_{\chi+\chi_0}(t_0)}\nonumber\\
	&= N e^{i\left(\phi_P(\ket{\psi_0(t_0)},\ket{\psi_{\chi_0}(t_0)})\right)}\int_0^{\chi_0}\,\d\chi\, \ket{\psi_{\chi}(t_0)}\nonumber\\
	&=e^{i\left[\phi_{\text{geom}}(\chi_0)+\chi_0\,(t_0-\langle\mathcal{\hat T}\rangle)\right]}\ket{\bar{\psi}_{\mathcal{\hat T},\,t_0}}.
\label{eqn:alternativeproofquantization1}
\end{align}
{{The only way the above equation is satisfied is if
\begin{align}
	&\ket{\bar{\psi}_{\mathcal{\hat T},\,t_0}}=0\label{eqn:inphasenotquantizedzero}\\
&~~~~\text{Or}~\nonumber\\
&\phi_P(\chi_0)=2\pi\mathbb{Z}\,.\nonumber\\ 
	\implies &\phi_{\text{geom}}(\chi_0)+\chi_0\,(t_0-\langle\mathcal{\hat T}\rangle)=2\pi \mathbb{Z}\,.
	\label{eqn:Quantizationarbitraryclosedorbit}
\end{align}
Thus, we have integral quantization of the Pancharatnam phase for the closed orbit.
However, the Pancharatnam phase is not a gauge invariant quantity.

To obtain a gauge invariant quantity, we consider in phase orbits, where
$\langle\mathcal{\hat T}\rangle=t_0$ (Eqn.~\eqref{eqn:localinphasesuperpositiongeneral}) holds.
Then, the condition~\eqref{eqn:Quantizationarbitraryclosedorbit} becomes 
\begin{align}
t_0=\langle\mathcal{\hat T}\rangle\implies \phi_{\text{geom}}(\chi_0)=2\pi\mathbb{Z}\,.
\label{eqn:Quantizationclosedorbitinphase}
\end{align}}}
Thus, horizontal superpositions on closed orbits give rise to exact integral quantization of geometric
phase.

Hence, for in phase orbits we have the following relations:
\begin{align}
&\text{Open orbits: }\ket{\bar{\psi}_{\mathcal{\hat T},\,\langle\mathcal{\hat{T}}\rangle}}=N_{\alpha_0,\,t_0}\int_{-\infty}^{\infty}\,\d\chi 
\,e^{i\chi(\langle\hat{\mathcal{T}}\rangle-\hat{\mathcal{T}})}\ket{\alpha_0}\nonumber\\
&\text{Closed Orbits: }\ket{\bar{\psi}_{\mathcal{\hat T},\,\langle\mathcal{\hat{T}}\rangle}}=N_{\alpha_0,\,t_0}\int_{0}^{\chi_0}\,\d\chi 
\,e^{i\chi(\langle\hat{\mathcal{T}}\rangle-\hat{\mathcal{T}})}\ket{\alpha_0}\nonumber\\
&\phi_{\text{geom}}(\chi_0)=2\pi\mathbb{Z},\qquad\text{else }\qquad\ket{\bar{\psi}_{\mathcal{\hat T},\,\langle\mathcal{\hat{T}}\rangle}}=0.
\label{eqn:inphaseGCS}
\end{align}

If $\Omega$ is K\"ahler, using Eqn.~\eqref{eqn:symplecticgeodesictriangle} for these privileged closed orbits, we get {\emph{exact}} Bohr-Sommerfeld 
quantization in phase space 
\begin{align}
\phi_{\text{geom}}(\chi_0)=2\pi\mathbb{Z}\implies	\mathscr{A}(\mathcal T,\alpha_0)=2\pi\mathbb{Z},
\label{eqn:generalBSquantization}
\end{align}
where $\mathscr{A}(\mathcal T,\alpha_0)$ is the symplectic area of the closed curve in $\Omega$ generated by $\mathcal{T}$ and passing through $\alpha_0$.
We will see illustrative examples in Sec.~\eqref{sec:examples}.

The $Q$ function (Eqn.~\eqref{eqn:qfunc}) of $\ket{\bar{\psi}_{\mathcal{\hat{T}},\,t_0}}$ is trivially invariant along each orbit
$\mathcal{O}_{\hat{\mathcal{T}},\,\alpha_0}$.
\begin{align}
e^{i\chi\hat{\mathcal{T}}}\ket{\bar{\psi}_{\mathcal{\hat{T}},\,t_0}}&=e^{it_0\chi}\ket{\bar{\psi}_{\mathcal{\hat{T}},\,t_0}}\nonumber\\
\bra{\bar{\psi}_{\mathcal{\hat{T}},\,t_0}}e^{-i\chi\hat{\mathcal{T}}}\ket{\alpha_0}&=
e^{-it_0\chi}\braket{\bar{\psi}_{\mathcal{\hat{T}},\,t_0}}{\alpha_0},\nonumber\\
\implies\rvert\braket{\bar{\psi}_{\mathcal{\hat{T}},\,t_0}}{e^{-i\chi\mathcal{T}}\alpha_0}\lvert^2&=\,
\rvert\braket{\bar{\psi}_{\mathcal{\hat{T}},\,t_0}}{\alpha_0}\lvert^2.
\label{eqn:Qfuncinvarintorbit}
\end{align}

We note that
integral Bohr-Sommerfeld quantization conditions have been considered in mathematical studies in the context of geometric quantization,
Refs.~\cite{Sniatycki_1980,Kirillov1990,Cushman_2013}. 
We hope that that this alternate approach proves to be useful and more approachable for physicists.
Also, we point out that Eqns.~\eqref{eqn:inphasenotquantizedzero},\eqref{eqn:Quantizationarbitraryclosedorbit},
\eqref{eqn:Quantizationclosedorbitinphase}
are valid for arbitrary $\Omega$, regardless of whether it is K\"ahler.

To recapitulate, we saw that an arbitrary eigenstate of $\mathcal{\hat T}\in\mathfrak{g}$ can be constructed as 
a $U(1)$ weighted average of CS over the group orbit $\mathcal{O}_{\mathcal{T},\,\alpha_0}$
(as long as $\alpha_0$ is not a fixed point of $\mathcal{T}$). However, for closed orbits, the sum 
vanishes unless the relative or Pancharatnam phase is quantized in integral multiples of $2\pi$.

For the eigenstate $\ket{\bar{\psi}_{\mathcal{\hat T},\,t_0}}$,
the orbit $\mathcal{O}_{\mathcal{T},\,\alpha_0}$ is privileged,
$t_0=\bra{\alpha_0}\mathcal{\hat{T}}\ket{\alpha_0}$. 
It allows for the expansion of $\ket{\bar{\psi}_{\mathcal{\hat T},\,t_0}}$ as an in phase superposition.
Further, the dynamical phase vanishes, and hence the geometric phase
(dependent solely on the orbit in $\Omega$ instead of the Hilbert space curve) is now quantized to $2\pi$ times an integer. What is even more striking 
is that an in phase superposition of the form of Eqn.~\eqref{eqn:averageclosedorbit} over an orbit where the geometric phase is not quantized vanishes.

We note that $\mathcal{O}_{\mathcal{T},\,\alpha_0}:\,t_0=\bra{\alpha_0}\mathcal{\hat{T}}\ket{\alpha_0}$ is not unique. 
See example in Sec.~\ref{subsec:2dsho}.

\section{Putting quantum states in phase space so that they stay put}
\subsection{Building up intuition: interference in phase space}\label{sec:intuition}
In the introduction, we motivated the use of the $Q$ function maximum to probe the location of states in phase space. 
Our interest, as evinced in the title is to put superpositions of coherent states along some curve in phase space
so that the resultant $Q$ function is maximized there.

We are concerned with 
the location of a continuous superposition of states along orbits as constructed in Section~\ref{subsec:constructiongeneral}. We remark that
this section is heuristic, it aims to outline the intuition which led us to the conjecture.

To develop some insight, first
consider the $Q$ function of the un-normalized superposition of two CS 
\begin{align}
\ket{\psi}=\ket{g_1}+e^{i\theta}\ket{g_2}.
\label{eqn:unnormalizedGCS}
\end{align}
Where is this superposition localized/$Q$ function maximized? 

\begin{align}
	&Q_{\ket{\psi}}(g)\propto |\braket{g}{g_1}+e^{i\theta}\braket{g}{g_2}|^2,\quad g\in\Omega,\nonumber\\
=&|\braket{g}{g_1}|^2+|\braket{g}{g_2}|^2+2\,\text{Re}\left[\frac{e^{i\theta}\braket{g}{g_2}\braket{g_1}{g}\braket{g_2}{g_1}}{\braket{g_2}{g_1}}\right]\nonumber\\
=& I_1(g)+I_2(g)+2\sqrt{I_1(g)I_2(g)}\cos\left(\theta_1+\theta_2\right).\label{eqn:GCSsuperposition1}\\
&I_i(g)=\vert\braket{g}{g_i}\vert^2;\,
\theta_1=\theta+\arg\,\braket{g_1}{g_2};\nonumber\\	
&\theta_2=\arg\,\mbox{Tr}\left[\hat{\rho}_{g_1}\hat{\rho}_{g}\hat{\rho}_{g_2}\right];\,\hat{\rho}_{g}=\ket{g}\bra{g}.\nonumber
\end{align}

Now, $Q_{\ket{\psi}}(g)$ in Eqn.~\eqref{eqn:GCSsuperposition1} can be rewritten as
\begin{align}
	Q_{\ket{\psi}}(g)&\propto I_1(g)+I_2(g)+2\sqrt{I_1(g)I_2(g)}\times\nonumber\\
	\cos&\left[\phi_P\left(\ket{g_1},e^{i\theta}\ket{g_2}\right)+\arg\Delta_3\,(\hat\rho_{g_1},\hat\rho_{g},\hat\rho_{g_2}) \right].
	\label{eqn:gcsconstructive}
\end{align}
It is useful to envision the $Q$ function in Eqn.~\eqref{eqn:gcsconstructive} as intensity resulting from ``interference" in phase space 
with two sources located at $g_1$ and $g_2$.

We make the physically plausible 
assumption that the $Q$ function of the CS state $\ket{g_i}$,
i.e. $I_i(g)$ in Eqn.~\eqref{eqn:GCSsuperposition1} is maximized 
at $g=g_i$. Also, in this subsection we restrict ourselves to Hermitian symmetric manifolds where Eqns.~\eqref{eqn:symplecticgeodesictriangle} and \eqref{eqn:BargmannInvariantGCS} 
hold. Hence, 
\begin{align}
	\label{eqn:geodesicareaBargmanninvariant}
\mathscr{A}(g_1,g_2,g_3)=\arg\Delta_3\,(\hat\rho_{g_1},\hat\rho_{g_2},\hat\rho_{g_3}),
\end{align}
the symplectic area of the geodesic triangle $(g_1,g_2,g_3)$ is equal to the argument of the 3-vertex Bargmann invariant.
Then, the lines of constructive interference in Eqn.~\eqref{eqn:gcsconstructive} are
\begin{align}
 \phi_P(\ket{g_1},e^{i\theta}\ket{g_2})+\mathscr{A}(g_1,g,g_2)=2\pi\mathbb{Z}.
\label{eqn:constructiveinterferencecondn}
\end{align}

{ {Thus, interference in phase space is driven by areas rather than propagation distances.}}
These ideas have been developed in more detail for SHO CS in ~Ref.\cite{khan2018geometric}.

Consider the scenario when the sources are in phase, i.e. $\phi_P(\ket{g_1},e^{i\theta}\ket{g_2})=0$. 
Then the geodesic joining $g_1$ and $g_2$ is a line of constructive interference,
and we expect $Q_{\ket{\psi}}(g)$ to fall off as we move away from this geodesic line.
When $\phi_P(\ket{g_1},e^{i\theta}\ket{g_2})\neq 0$,
the location of $\ket{\psi}$ is shifted in the perpendicular direction.


Carrying over these ideas to continuous superpositions, we expect that at least for open curves, putting an in phase superposition
along a geodesic results in constructive interference along the said line of superposition. 
Using Eqn.~\eqref{eqn:constructiveinterferencecondn} and the fact that $\mathscr{A}(g_1,g,g_2)$ vanishes if $g$ lies on the geodesic joining
$g_1$ and $g_2$, any two in phase states interfere constructively on any other point on the geodesic.
We will see examples of this
in Eqns.~\eqref{eqn:posinphase},~\eqref{eqn:mominphase}.

Another piece of evidence in anticipation of the conjecture to be presented is squeezing for in phase superposition of two SHO CS 
in a direction perpendicular to the line joining them.~\cite{Janzky1,khan2018geometric}. Indeed, as more states 
are added to this superposition, squeezing becomes more pronounced. Our construction takes this to the limit by 
considering a continuous superposition.

Of course, these arguments are heuristic and do not constitute a proof. The $n$th line of
constructive interference is generally not even a bright fringe as $I_i(g)$ falls off as we move away from
$g_i$.

Further, it turns out constructive interference can happen along curves which are not geodesics
(e.g. circles for the Fock state).
%
%
Non-local effects like the net geometric
phase over a closed orbit also play an important role as we have already seen.
The general case turns out to be a complex 
multi source interference effect in phase space driven by a combination of Bargmann and Pancharatnam phase, 
and we do not attempt to solve it.

Guided by these observations and buoyed by the examples (to be presented later in Sec.~\ref{sec:examples}), now 
we make a conjecture about the eigenstates of the generator.
\subsection{Conjecture about the \texorpdfstring{$Q$}{Q} function: location of the eigenstates 
\texorpdfstring{$\ket{\bar{\psi}_{\mathcal{\hat T},t_0}}$}{} in \texorpdfstring{$\Omega$}{omega}}
Consider the coherent state system $\{G,T,\ket{0}\}$ of the Lie group $G$ 
built from the fiducial state vector $\ket{0}$ with
the largest isotropy subalgebra. (refer to~\cite{CoherentPerelomov} for details) $H$ is the isotropy subgroup of the
fiducial state. The coherent space manifold $\Omega=G/H$. The group $G$ acts transitively on $\Omega$.
We assume that $\Omega$ is K\"ahler.

Let $\mathcal{T}$ be a generator of the group $G$, i.e. 
$\mathcal{T}\in\mathfrak{g}$, the Lie algebra and is Hermitian.
We denote by $\ket{\bar{\psi}_{\mathcal{\hat T},\,t_0}}$ an eigenstate of $\mathcal{\hat{T}}$
with eigenvalue $t_0$. 
The $Q$ function of $\ket{\bar{\psi}_{\mathcal{\hat T},\,t_0}}$, $Q_{\ket{\bar{\psi}_{\mathcal{\hat T},\,t_0}}}$
is constant along each orbit $\mathcal{O}_{\mathcal{T},\,\alpha}$ of $\mathcal{\hat{T}}$ in $\Omega$.

We conjecture that the maximum value of $Q$ occurs along a curve in $\Omega$ which is an orbit
of $\mathcal{T}$, $\mathcal{O}_{\mathcal{T},\alpha_0}$, such that
$\langle\mathcal{\hat T}\rangle=\bra{\alpha_0}\mathcal{\hat T}\ket{\alpha_0}=t_0$.
For closed orbits, we 
also demand that the symplectic area enclosed is an integral multiple of $2\pi$.\par

If the eigenspace associated with $t_0$ is non-degenerate, there is a unique orbit $\mathcal{O}_{\mathcal{T},\,\alpha_0}$
associated with the eigenstate. On the other hand, with degenerate eigenspaces, one expects to find an infinite number of 
disjoint orbits satisfying
$\langle\mathcal{\hat T}\rangle=t_0$.(see Section~\ref{subsec:2dsho} for an example)
An average of the form of Eqns.~\eqref{eqn:generalopenorbitaverage},\eqref{eqn:averageclosedorbit} over the orbit $\mathcal{O}_{\mathcal{T},\,\alpha_0}$
generates an eigenstate whose $Q$ function is maximized over $\mathcal{O}_{\mathcal{T},\,\alpha_0}$.

We can express $\ket{\bar{\psi}_{\hat{\mathcal{T}},\,t_0}}$  as a CS expansion over another orbit $\mathcal{O}_{\mathcal{T},\,\alpha'_0}$,
but the state as diagnosed by the $Q$ function maximum still lies at $\mathcal{O}_{\mathcal{\hat{T}},\,\alpha_0}$.
The coherent state expansions in
Eqn.~\eqref{eqn:inphaseGCS} put 
along in phase orbits (which are their curves of $Q$ function maximum) stay there.


We show the validity of this conjecture in illustrative examples in the following section.
In phase expansions along 
$\mathcal{O}_{\mathcal{T},\,\alpha_0}$ are tantamount to constructive interference along the line of superposition.

It is important to note that other restrictions might be necessary.
In particular, all the examples we consider are Hermitian symmetric manifolds, where Eqn.~\eqref{eqn:geodesicareaBargmanninvariant} holds true. 
Hence, one line of inquiry would be to consider a 
manifold which is not Hermitian symmetric. 
However, mostly generators 
produce orbits which are not geodesics, and hence we do not expect 
Eqn.~\eqref{eqn:geodesicareaBargmanninvariant} to help with constructive interference. Thus, we have assumed the validity 
of the conjecture in general for K\"ahler manifolds. 

In any case, it would be good to have a proof/disproof of this conjecture or example 
where it fails. However, even if the general conjecture as stated fails, considering its validity in the number of cases shown below, we expect some 
structure present which should
hold true with some restrictions.

For a horizontal closed curve, the relative (and geometric phase) is determined by area enclosed in (K\"{a}hler) phase space.
The $Q$ function is maximized along the curve as well.
These facts 
have been exploited for SHO CS~\cite{khan2018geometric} to understand the emergence of areas 
while calculating the interference term in 
inner products between states whose representations
as line integral horizontal superpositions intersect. 
This phenomenon is popularly known as interference in phase space \cite{Wheeler_1985,Schleich_1987}.
Our construction in Sec.~\ref{subsec:constructiongeneral} which provides a recipe for obtaining eigenstates as horizontal superpositions
over orbits shows that this scheme can be generalized to other CS systems as well. 
We will show next that this conjecture is true for a variety of groups, $H_4$, $SU(2)$ and $SU(1,1)$.
Indeed interference in phase space
has been studied for spherical phase space~\cite{LassigMilburn} (associated with $SU(2)$),
and hyperbolic phase space~\cite{Chaturvedi_Milburn} ($SU(1,1)$).

\section{Examples}\label{sec:examples}
\subsection{Simple Harmonic Oscillator coherent states}\label{subsec:SHOcoherentstates}
We apply the general construction from Section~\ref{subsec:constructiongeneral} to simple harmonic oscillator (SHO) 
coherent states. 
These results are there in~\cite{khan2018geometric}, but we now present them from a group theoretic perspective.
The relevant group $H_4$ is generated by the algebra $h_4$ \cite{CoherentPerelomov,GilmoreCoherentReview} with the 
generators $\{\mathbbm{1},\hat{q},\hat{p},\hat{n}=\hat{a}^\dagger\hat{a}\}$, where the 
annihilation operator $\hat{a}=\frac{1}{\sqrt{2}}(\hat{q}+i\hat{p})$. 
The Hilbert space is spanned
by the eigenkets of the number operator $\hat{n}\ket{n}=n\ket{n}$.

Any element of the $H_4$ oscillator group takes the form
$e^{{i(p\hat{q}-q\hat{p})}}e^{{i(t\mathbbm{1}+y\hat{n})}}$. 
We choose the harmonic oscillator ground state as the fiducial state $\ket{0}$. It satisfies $\hat{n}\ket{0}=0$, and
has the maximal isotropy subalgebra $\{\mathbbm{1},\hat a,\hat{n}\}$.

The maximal isotropy group of $\ket{0}$ consists of the elements
$\exp[{i(t\mathbbm{1}+y\hat{n})}]$. 
$\exp[{i(t\mathbbm{1}+y\hat{n})}]\ket{0}=e^{it}\ket{0}.$

Thus, the CS are indexed by points $(q,p)$ on the two dimensional real phase plane,
$(q,p)\in \mathbb{R}^2$ or the complex combination 
$z=\frac{1}{\sqrt{2}}(q+ip);\,z\in\mathbb{C}$. 
$\ket{q,p}$ is obtained by the action of the 
displacement operator $D(q,p)$ on $\ket{0}$.
The various equivalent definitions of CS we shall use are
\begin{align}
 \ket{q,p}\equiv\ket{z}&=D(q,p)\ket{0}=e^{i(p\hat{q}-q\hat{p})}\ket{0},\nonumber\\
 \ket{z}=D(z)\ket{0}&=e^{z\hat{a}^\dagger-\bar{z}\hat{a}}\ket{0}=
 e^{-\vert z\vert^2/2}e^{z\hat{a}^{\dagger}}\ket{0}.
\end{align}
The canonical CS are eigenstates of the annihilation operator $\hat{a}\ket{z}=z\ket{z}$.

We first obtain expressions for eigenstates of $\hat{q},\hat{p}$ which generate translations
in $p$ and $q$ respectively, and have open orbits in phase space.
\begin{align*}
e^{ip\hat{q}}\ket{q_0,p_0}=e^{\frac{i}{2}q_0p} \ket{q_0,p_0+p};\,\bra{q_0,p_0}\hat{q}\ket{q_0,p_0}=q_0.
\end{align*}
$\mathcal{O}_{q,\,(q_0,0)}$ (where $\alpha_0=(q_0,0)$ in the notation of the previous section)
is the line $q=q_0$ parallel to the p-axis.
The position eigenstate\,$\ket{q_1;\text{pos}}$ by Eqn.~\eqref{eqn:generalopenorbitaverage} is
\begin{align*}
	\ket{q_1;\text{pos}}\equiv\ket{\bar{\psi}_{\hat{q}\,,q_1}}&=N\int_{-\infty}^{\infty}\, \d p\, e^{i(q_1 -{q_0}/{2}) p}\ket{q_0,-p}\\
&=N\int_{-\infty}^{\infty}\, \d p\, e^{ip\,({q_0}/{2}-q_1)}\ket{q_0,p}.
\end{align*}
Calculating the inner product 
with another position eigenstate
$\bra{x_0;\text{pos}}q_0;\text{pos}\rangle=\delta(x_0-q_0)$ and using
\[\braket{q',\text{pos}}{q,p} = \frac{1}{\pi^{1/4}}\exp[-\frac{1}{2}{(q'-q)}^2 + i p(q'-q/2)],\]
yields the normalization $N$.
Finally, we get
\begin{align}
	\ket{q_1;\text{pos}}&=\frac{e^{\frac{1}{2}(q_0-q_1)^2}}{2\pi^{3/4}}\int_{-\infty}^{\infty} \d p\, e^{ip(\frac{q_0}{2}-q_1)}\ket{q_0,p}.
	\label{eqn:positionsuperpositionarbitrary}
\end{align}
However, this is not an in phase superposition. As was shown in Section~\ref{subsec:constructiongeneral}, this will
be so for $\ket{q_1;\text{pos}}$ when we choose the orbit such that $\langle\hat q\rangle=q_1$. This restricts us to the orbit 
$\mathcal{O}_{q,\,(q_1,0)}$.  Putting $q_0=q_1$ in Eqn.~\eqref{eqn:positionsuperpositionarbitrary} leads us to the in phase expansion
\begin{align}
\ket{q_1;\text{pos}}&=\frac{1}{2\pi^{3/4}}\int_{-\infty}^{\infty} \d p\, e^{-\frac{i}{2}q_1p}\ket{q_1,p}.  
\label{eqn:posinphase}
\end{align}

A completely analogous procedure for $\hat p$, 
yields 
\begin{align}
\ket{p_1;\text{mom}}&=\frac{e^{\frac{1}{2}(p_0-p_1)^2}}{2\pi^{3/4}}\int_{-\infty}^{\infty} \d q\, e^{iq\left(p_1-\frac{p_0}{2}\right)}\ket{q,p_0}. 
\label{eqn:momentumarbitrarysuperposition}	
\end{align}
Restricting ourselves to the orbit $\mathcal{O}_{p,\,(p_1,0)}$, we obtain the in phase expansion
\begin{align}
\ket{p_1;\text{mom}}&=\frac{1}{2\pi^{3/4}}\int_{-\infty}^{\infty} \d q\, e^{\frac{i}{2}qp_1}\ket{q,p_1}.  
\label{eqn:mominphase}
\end{align}
The in phase superpositions (Eqns.~\eqref{eqn:posinphase} and \eqref{eqn:mominphase}) for position and momentum eigenstates are along geodesics
in $\mathbb{R}^2$. The symplectic area, and hence the argument of the Bargmann invariant vanishes for any triad of 
states on the orbit. Hence they are not just local, but {{global}} in phase expansions.
For the position and momentum eigenstate expansions (Eqns.~\eqref{eqn:positionsuperpositionarbitrary} and ~\eqref{eqn:momentumarbitrarysuperposition}),
we see that the superposition coefficients increase exponentially the further we move away from the in phase superpositions at $q=q_1$ and $p=p_1$
respectively, exhibiting efficient superposition along the in phase orbits as alluded to in the introduction. 

Now, we turn to $\hat{n}$ which generates closed orbits (circles) in phase space.
\begin{align}
e^{-i\theta\hat{n}}\ket{z}=\ket{z\,e^{-i\theta}};\,\,\bra{z}\hat{n}\ket{z}=\vert z\vert^2. 
\end{align}
As we saw in Section~\ref{subsec:constructiongeneral}, geometric phase considerations now become important.
Let us consider CS along a closed curve in $\Omega=\mathbb{R}^2$,
\begin{align*}
 \mathscr{C}=\{\ket{z(\chi)}:\,\ket{z(0)}=\ket{z(\chi_0)}\}.
\end{align*}
Since $\phi_{\text{geom}}(\chi_0)$ is invariant under $U(1)$ transformations, we work with the
horizontal curve (Eqn.~\eqref{eqn:horizontalcurvegeometricphase}) 
\begin{align*}
	\mathscr{C}_{\text{hor}}&=\{\ket{z'(\chi)}=\ket{z(\chi)}\,e^{i\phi(\chi)}\},\\
	\phi(\chi)&=i\int_0^{\chi}\d\chi_1\bra{z(\chi_1)}\frac{d}{d\chi_1}\ket{z(\chi_1)}.
\end{align*}
The dynamical phase vanishes, and $\phi_{\text{geom}}(\chi_0)$ 
is simply obtained by the Pancharatnam phase accumulated, $\text{arg }\braket{z'(0)}{z'(\chi_0)}$. 
\begin{align}
	&\phi_{\text{geom}}(\chi_0)=\text{arg}\braket{z'(0)}{z'(\chi_0)}=i\int_0^{\chi_0}\d\chi\bra{z(\chi)}\frac{\text{d}}{\d\chi}\ket{z(\chi)}\nonumber\\
	&=i\int_0^{\chi_0}\d\chi\frac{-1}{2}\left(\bar{z} \frac{\text{d}z}{\d\chi}+z\frac{\text{d}\bar{z}}{\d\chi}\right)+\bra{z}\hat{a}^\dagger\ket{z}
	\frac{\text{d}z}{\d\chi}\nonumber\\
	&=\frac{i}{2}\int_0^{\chi_0}\d\chi\left(\bar{z} \frac{\text{d}z}{\d\chi}-z\frac{\text{d}\bar{z}}{\d\chi}\right) =\frac{1}{2}\int_0^{\chi_0}(q\d p-p\d q)
	\nonumber\\
	&=-\oiint_{S(\mathscr{C})}\d p\wedge \d q.\label{eqn:berrycanonicalclosed}
\end{align}
Thus, the geometric phase for a closed curve in $\Omega$ is given by the negative of the symplectic area enclosed.
(Area enclosed is taken to be positive when we traverse the curve in the anticlockwise direction)

Using Eqn.~\eqref{eqn:averageclosedorbit} 
\begin{align*}
	&\ket{n_0}\equiv\ket{\bar{\psi}_{\hat{n},n_0}}= \mathcal{N}_{n_0}\int_0^{2\pi} d\theta\, 
e^{in_0\theta}\,\ket{\frac{r_0}{\sqrt{2}}e^{-i\theta}},\\
&\ket{r_0\cos\theta,\,r_0\sin\theta}\equiv\ket{\frac{r_0}{\sqrt{2}}e^{i\theta}}.
\end{align*}
We have chosen as the orbit $\mathcal{O}_{n,(r_0,0)}$, which is a circle of radius $r_0$ centred at the origin.
Positivity of $\hat n$ ensures $n_0\geq 0$.
On $\mathcal{O}_{n,(r_0,0)}$, $\langle\hat n\rangle=\frac{r_0^2}{2}$.
Since the curve is traversed clockwise, Eqns.~\eqref{eqn:dynphasealongorbit} and \eqref{eqn:berrycanonicalclosed}, give
\begin{align*}
&\phi_{\text{geom}}=\pi r_0^2;\,\phi_{\text{dyn}}=2\pi \left(n_0-\frac{r_0^2}{2}\right);\\
&\phi_{\text{geom}}\,+\,\phi_{\text{dyn}}=2\pi\,n_0.
\end{align*}
But by Eqn.~\eqref{eqn:Quantizationarbitraryclosedorbit} for closed orbits,
\begin{align*}
	2\pi n_0=2\pi\mathbb{Z}\,;~\because n_0\geq 0\implies n_0\in \mathbb{N}\cup\{0\}.
\end{align*}
Thus, we get the quantization of the eigenvalues of $\hat{n}$, which 
is consistent.

The normalization $\mathcal{N}_m$ for the Fock state is determined by calculating $\braket{m}{p}=\delta_{mp}$,
where $\ket{p}$ is another Fock state 
\begin{align}
&\ket{m}=\frac{\sqrt{m!}}{2\pi}\left(\frac{\sqrt{2}}{r_0}\right)^m e^{\frac{r_0^2}{4}}\int_0^{2\pi}\d\theta\,e^{im\theta}\,
\ket{\frac{r_0\,e^{-i\theta}}{\sqrt{2}}}.
\label{eqn:Fockoverarbitrarycircle}
\end{align}
As expected, $\mathcal{N}_m=\dfrac{\sqrt{m!}}{2\pi}\left(\dfrac{\sqrt{2}}{r_0}\right)^m e^{\frac{r_0^2}{4}}$ is minimized at $r_0=\sqrt{2m}$ for
the Fock state $\ket{m}$.

For in phase superpositions, $\langle\hat n\rangle=\frac{r_0^2}{2}=m$ (by Eqn.~\eqref{eqn:localinphasesuperpositiongeneral}),
thus specifying the orbit $r_0=\sqrt{2m}$.
\begin{align}
	&\ket{m}=\frac{\sqrt{m!}}{2\pi}m^{-\frac{m}{2}} e^{\frac{m}{2}}\int_0^{2\pi} \d\theta\,e^{im\theta}\,
\ket{\sqrt{m}\,e^{-i\theta}}.
\label{eqn:Fockinphase}
\end{align}
Over orbits of radius $\sqrt{2m}$, Eqn.~\eqref{eqn:berrycanonicalclosed} implies exact Bohr-Sommerfeld quantization for the Fock state $\ket{m}$,
\begin{align}
 \oint pdq=2\pi m.
\label{eqn:BSSHOstates}
\end{align}
Alternatively consider
\begin{align}
	\ket{\bar\psi}=\mathcal{N}\int_0^{2\pi}\d\theta\,e^{i\frac{r_0^2}{2}\theta}
\,\ket{\frac{r_0}{\sqrt{2}}\,e^{-i\theta}};\quad r_0^2\neq 2\mathbb{N}.
\end{align}
This is an in phase superposition. Further,
it is easily checked that $\braket{m}{\bar\psi}=0\,\forall m$, which imples $\ket{\bar\psi}=0$.
This is to be expected from Eqns.~\eqref{eqn:inphasenotquantizedzero} and~\eqref{eqn:Quantizationclosedorbitinphase}  as $\phi_{\text{geom}}$
is not an integral multiple of $2\pi$.

Finally, we look at the $Q$ distribution of the eigenstates of $\hat{q},\hat{p}$ and $\hat{n}$.
$Q_{\ket{\psi}}=\dfrac{1}{2\pi}\vert\braket{q,p}{\psi}\vert^2$.
\begin{figure*}[htbp!]
\centering
\subfigure[t][{\label{fig:Ellipse}(colour online) Heat map of the $Q$ function of in phase superposition of states along the ellipse $\frac{q^2}{4}+p^2=1$.
The location of the states are denoted by red dots. The plot clearly indicates that the maximum of the $Q$ function does not lie along the 
superposition.
}]{ \includegraphics[width=0.46\textwidth]{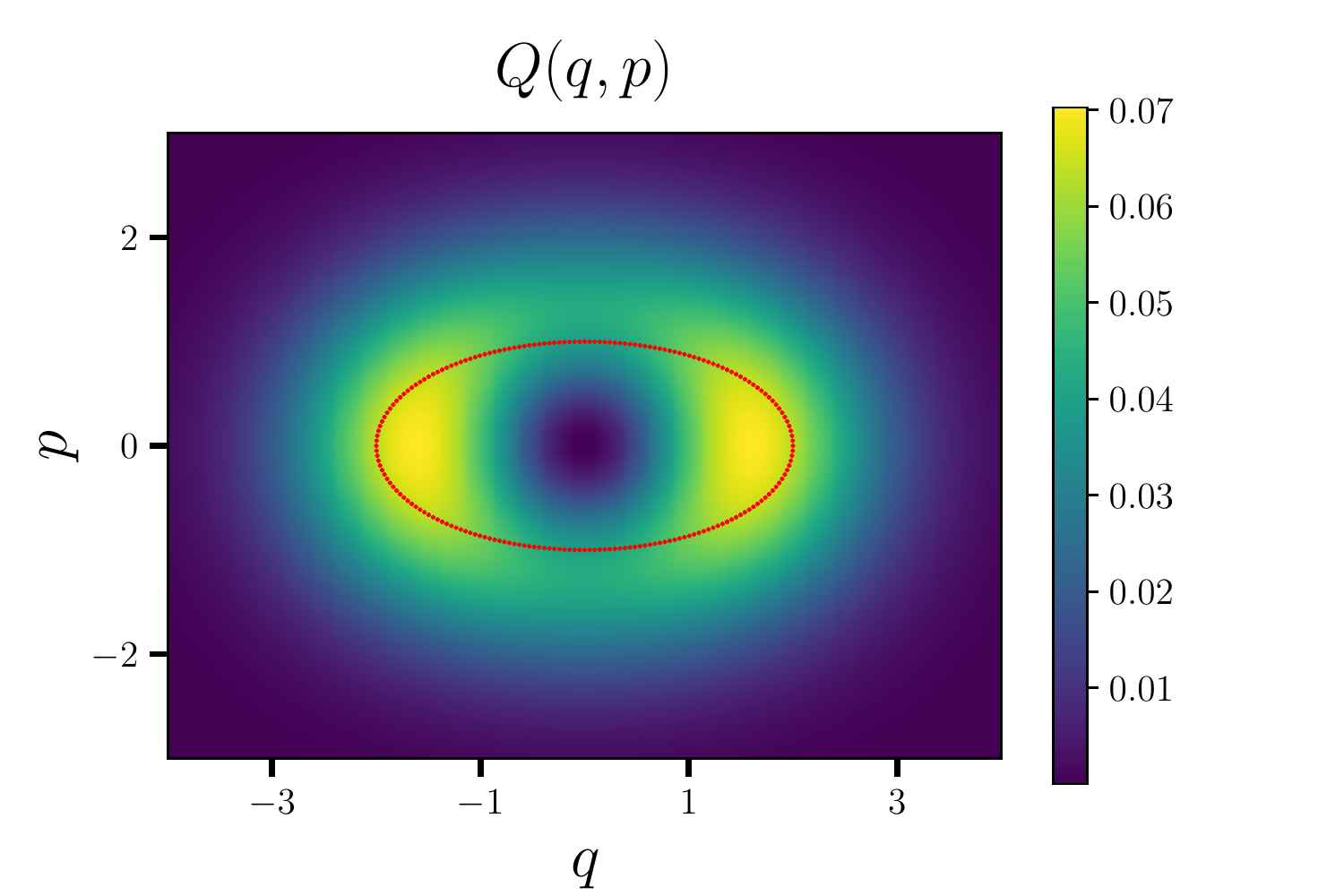}}\hfill
\subfigure[t][{\label{fig:Ellipseqvalalongp}(colour online) Variation of the $Q$ function of Fig.~\ref{fig:Ellipse} 
along the major axis $p=0$. The solid black lines denote the points on the
ellipse $(\pm 2,0)$, the dashed blue lines denote the position of the maximum at $(\pm 1.613,0)$.
}]{ \includegraphics[width=0.46\textwidth]{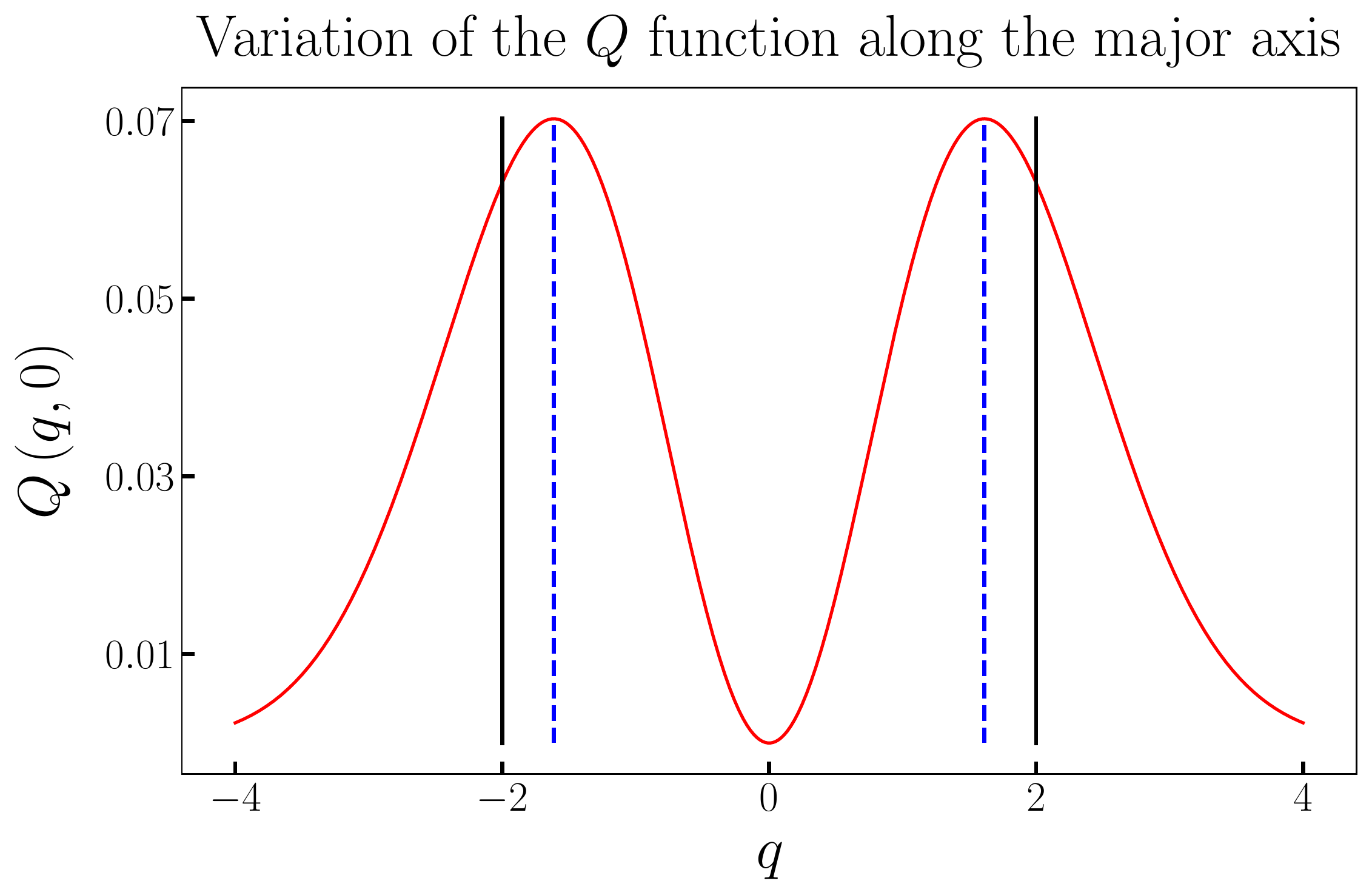}}
\caption{\label{fig:Driven Ellipse}	
Analysis of the $Q$ function of an in phase superposition of CS along the ellipse $\frac{q^2}{4}+p^2=1$}
\end{figure*}
\begin{align}
\label{eqn:QdistributionH4}
Q_{\ket{q_1,\text{pos}}}(q,p)&=\frac{1}{2\pi^{3/2}}\exp\left[-(q-q_1)^2\right];\\
	Q_{\ket{p_1,\text{mom}}}(q,p)&=\frac{1}{2\pi^{3/2}}\exp\left[-(p-p_1)^2\right];\nonumber\\	
	Q_{\ket{m}}(q,p)=\frac{1}{2\pi m!}&\left[\frac{p^2+q^2}{2}\right]^m \nonumber
	\exp\left[-\frac{1}{2}(p^2+q^2)\right].
\end{align}
We have included the usual factor of $2\pi$ coming from the 
measure $\dfrac{dqdp}{2\pi}$ in comparison with Eqn.~\eqref{eqn:qfunc}.

The $Q$ distributions are indeed maximized along the privileged orbits of their in phase superpositions,
$q=q_1$, $p=p_1$ and $r^2=2m$ as we have conjectured.

There is no obvious generator in $h_4$ which has elliptical orbits.
As a prelude to Section~\ref{subsec:coherentstatesellipse}, we now ask what happens if we nevertheless put a continuum of in phase states on an ellipse.
Does the resultant sum have a $Q$ function maximum along the in phase superposition? We take a numerical approach
to answer the question. We put $300$ equispaced states on the ellipse $\frac{q^2}{4}+p^2=1$, such that the neighbouring states are in phase with each other.
The enclosed area is $2\pi$, thus ensuring that the beginning and the penultimate states are in phase with each other as well.

It is quite obvious from Fig.~\ref{fig:Ellipse} that the maximum of the $Q$ function does not lie on the ellipse of
superposition. In Fig.~\ref{fig:Ellipseqvalalongp} we plot the variation of the $Q$ function as we move along the major axis through the origin,
and we see more concretely that this is indeed the case.
\subsection{Simple harmonic oscillator coherent states on an ellipse}\label{subsec:coherentstatesellipse}
In Section~\ref{subsec:SHOcoherentstates}, we briefly considered the difficulty of putting states on an ellipse
such that they stayed there, or in other words, such that the $Q$ function was maximized along the superposition.

However, there clearly is a quadratic operator whose orbit is an ellipse.
Let us consider $\hat{n}_s$.
\begin{align}
	&\hat{n}_s=\hat{a}_s^\dagger\hat{a}_s,~\hat{a}_s=\frac{1}{\sqrt{2}}(s\hat{q}+i\frac{\hat{p}}{s}),~[a_s,a_s^\dagger]=1.\nonumber\\
&e^{i\theta\hat{n}_s}D(q,p)e^{-i\theta\hat{n}_s}=D(q\cos\theta-\frac{p}{s^2}\sin\theta,p\cos\theta+qs^2\sin\theta)\nonumber\\
&\implies e^{i\theta\hat{n}_s}D(\frac{c}{s},0)e^{-i\theta\hat{n}_s}=D(\frac{c}{s}\cos\theta,cs\sin\theta).
\label{eqn:operatordrivealongellipse}
\end{align}
In this case, we have chosen the operators $\hat{a}_s$ so as to introduce a natural scale into the problem and reciprocal scaling 
of the variables $\hat{q}$ and $\hat{p}$. 
However, $\hat{n}_s\notin h_4$, and thus the validity of the conjecture is not violated.

We therefore construct the CS system for the slightly modified algebra $h_{4,s}=\{\mathbbm{1},\hat{q},\hat{p},\hat{n}_s\}$ 
and determine whether the eigenstates of $\hat{n}_s$ are maximized along an ellipse.
The new fiducial state
$\ket{0}_s$ obeys $\hat{a}_s/\hat{n}_s\ket{0}_s=0$. As expected, $\ket{0}_s$ is just a scaled version of the ordinary SHO ground state.
\[\braket{q;\text{ pos}}{0}_s=\frac{\sqrt{s}}{\pi^{1/4}}e^{-\frac{s^2q^2}{2}};
~~\braket{p;\text{ pos}}{0}_s=\frac{1}{\sqrt{s}\pi^{1/4}}e^{-\frac{p^2}{2s^2}}.\]
\begin{align}
\label{eqn:definingscaledSHO}
&	\ket{q,p}_s\equiv\ket{z}_s=D(q,p)\ket{0}_s=e^{i(p\hat q-q\hat p)}\ket{0}_s,\\
&	\ket{z}_s=D(z)\ket{0}_s=e^{z\hat{a}_s^\dagger-\bar{z} a_s}\ket{0}_s,~z=\frac{1}{\sqrt{2}}\left(qs+i\frac{p}{s}\right).\nonumber
\end{align}
The new Fock states are
\begin{align*}
	\ket{m}_s=\frac{\left(\hat{a}_s^\dagger\right)^m}{\sqrt{m!}}\ket{0}_s,\quad \hat{m}_s\ket{m}_s=m\ket{m}_s.
\end{align*}
We focus on the eigenstates generated by $\hat{n}_s$. Consider the orbit $\mathcal{O}_{n_s,\,(a,0)}$.
The classical orbit in $\Omega$ is 
\[\frac{q^2}{a^2}+\frac{p^2}{a^2s^4}=1.\] 
The expansion of $\ket{m}_s$ along $\mathcal{O}_{n_s,\,(a,0)}$
is 
\begin{align*}
	\ket{m}_s=\mathcal{N}_m\int_0^{2\pi}\d\theta\,e^{im\theta}\ket{a\cos\theta,-as^2\sin\theta}_s.
\end{align*}
We check that Eqn.~\eqref{eqn:Quantizationarbitraryclosedorbit} is satisfied
\begin{align*}
\label{eqn:quantizationarbitraryorbitellipse}
	\phi_{\text{geom}}=\pi a^2s^2;\quad&\langle\hat{n}_s\rangle=\vert z\vert^2=\frac{a^2s^2}{2};\nonumber\\
	\phi_{\text{geom}} +2\pi\,(m-\langle\hat{n}_s\rangle)&=\pi a^2s^2+2\pi(m-\frac{1}{2}a^2s^2)=2\pi m,\\
	m\,\in &\, \mathbb{N}\,\cup \{0\}.
\end{align*}
Including the normalization $\mathcal{N}_m$,
\begin{align}
	\ket{m}_s=e^{\frac{a^2s^2}{4}}\frac{2^{m/2}\sqrt{m!}}{a^ms^m}\int_0^{2\pi}\frac{\d\theta}{2\pi}e^{im\theta}\ket{a\cos\theta,-as^2\sin\theta}_s.
\end{align}
In phase superposition happens when \[\langle\hat{n_s}\rangle=\frac{1}{2}a^2s^2=m\implies a=\frac{\sqrt{2m}}{s}.\]
Over the privileged orbits $\mathcal{O}_{{n}_s,\,(\frac{\sqrt{2m}}{s},\,0)}$ we get the in phase superposition
\begin{align}
	\ket{m}_s=\frac{e^{m/2}}{m^{m/2}}\frac{\sqrt{m!}}{2\pi}\int_0^{2\pi}{\d\theta}e^{im\theta}\ket{\frac{\sqrt{2m}}{s}\cos\theta,-s\sqrt{2m}\sin\theta}_s.
\end{align}
$\phi_{\text{geom}}=2\pi m$ for these orbits.

We see that the $Q$ function of the eigenstate $\ket{n}_s$ is
\begin{align}
	Q_{\ket{n}_s}= \frac{1}{2\pi}\frac{\left(q^2s^2 + \frac{p^2}{s^2}\right)^n}{n!~2^n} \exp\left[-\frac{1}{2}\left(q^2s^2+\frac{p^2}{s^2}\right)\right]. 
\end{align}
Indeed, the $Q$ function is maximized along the ellipse $\left(q^2s^2+\frac{p^2}{s^2}\right)=2n$. For the state $\ket{1}_s$ and $s=\frac{1}{\sqrt{2}}$,
the maximum lies along  $\left(\frac{q^2}{4}+{p^2}\right)=1$.
\subsection{2d isotropic simple harmonic oscillator: dealing with degeneracy}\label{subsec:2dsho}
In this section, we consider the 2d isotropic SHO. It closely follows the 1d SHO from Sec.~\ref{subsec:SHOcoherentstates}, 
We introduce it only to briefly highlight aspects of degeneracy that were absent in the 1d case.

The Lie algebra $\mathfrak{g}=\{\mathbbm{1},\hat{q}_1,\hat{p}_1,\hat{q}_2,\hat{p}_2,\hat{n}=\hat{a}_1^{\dagger}\hat{a}_1+\hat{a}_2^{\dagger}\hat{a}_2\}$,
where $\hat{a}_i=\frac{1}{\sqrt{2}}\left(\hat{q}_i+i\hat{p}_i\right)$. Points in the CS 
manifold are now given by $(q_1,p_1,q_2,p_2)$.

The eigenspace of $\hat{n}$ with eigenvalue $1$ is 2 dimensional. In the basis $\ket{n_1,n_2}$ (the simultaneous
eigenvectors of $\hat{n}_1=\hat{a}^\dagger_1\hat{a}_1$ and $\hat{n}_2=\hat{a}^\dagger_2\hat{a}_2$) it is spanned by the states
$\ket{0,1}$ and $\ket{1,0}$.

By Eqn.~\eqref{eqn:localinphasesuperpositiongeneral}, in phase superpositions $\ket{\bar\psi_{\hat{n},1}}$ must
be over orbits such that
\[q_1^2+p_1^2+q_2^2+p_2^2=2.\]
Being a degenerate system, one can find infinitely many disjoint orbits which satisfy the above criterion.
For example, consider the the family of orbits indexed by $\alpha$, $\mathcal{O}_{\hat{n},\,(\sqrt{2}\cos\alpha,0,\sqrt{2}\sin\alpha,0)}$. 
Each orbit is given as 
\[\{(\sqrt{2}\cos\alpha\cos\theta,\sqrt{2}\cos\alpha\sin\theta,\sqrt{2}\sin\alpha\cos\theta,\sqrt{2}\sin\alpha\sin\theta)\},\]
where $\theta\in[0,2\pi)$.

We conclude by giving the basis states $\ket{1,0}$ and $\ket{0,1}$ as in phase superpositions.

The ket $\ket{1,0}$ can be expanded over the orbit $\mathcal{O}_{\hat{n},\,(\sqrt{2},0,0,0)}$ as
\begin{align}
	\ket{1,0}=\mathcal{N}\int_0^{2\pi}\d\,\theta\,\,e^{i\theta}\,\ket{\sqrt{2}\cos\theta,-\sqrt{2}\sin\theta,0,0}. 
\label{eqn:ket10}
\end{align}
Similarly, $\ket{0,1}$ expanded over the orbit $\mathcal{O}_{\hat{n},\,(0,0,\sqrt{2},0)}$ is
\begin{align}
	\ket{0,1}=\mathcal{N}\int_0^{2\pi}\d\,\theta\,\,e^{i\theta}\,\ket{0,0,\sqrt{2}\cos\theta,-\sqrt{2}\sin\theta}. 
\label{eqn:ket01}
\end{align}

\subsection{\texorpdfstring{$su(2)$}{su(2)} coherent states}
The group $SU(2)$ consists of the set of $2\times 2$ unitary matrices with determinant unity. The general element of
$SU(2)$ can be represented as $a^0\mathbbm{1}_{2\times 2}+ia^1\sigma_x+ia^2\sigma_y+ia^3\sigma_z$, 
$\sum\limits_{i=0}^3 (a^i)^2=1,\,a^i\in\mathbb{R}$, where $\sigma_i$ are the Pauli spin matrices.
Hence, the group manifold of $SU(2)$ can be identified with the 3 sphere $S^3$.

In general, $SU(2)$ admits $(2j+1)$ dimensional spin $j$ representations,  
$j\in \mathbb{N}/2$. 
The generators obey the relations $[\hat{J}_i,\hat{J}_j]=i\epsilon_{ijk}\hat{J}_k$, $i,j,k\in x,y,z$. 
(the spin $1/2$ generators are $\hat{J}_i=\frac{\sigma_i}{2}$)
The corresponding
raising and lowering operators are $\hat{J}_\pm=\hat{J}_x \pm i\hat{J}_y$. The set of states in the Hilbert space
$\mathcal{H}$ are labelled by the simultaneous eigenstate of $\hat{J}^2$ and $\hat{J}_z$, $\ket{j,m}$.

$SU(2)$ CS have been introduced and studied in detail in  
\cite{radcliffe1971,atkins1971angular,arecchi1972atomic,perelomov1972coherent}
We choose as the fiducial state, $\ket{0}=\ket{j,j}$, which has the maximal isotropy subalgebra 
$\{\mathbbm{1},\hat{J}_z,\hat{J}_+\}$. The isotropy subgroup $H$ of $\ket{j,j}$ is $e^{i\theta\hat{J}_z}$,
$e^{i\theta\hat{J}_z}\ket{j,j}=e^{i\theta j}\ket{j,j}$.

 Any group element of $SU(2)$ can be represented as 
$e^{\eta_-\hat{J}_-}e^{2\eta_0 \hat{J}_z}e^{\eta_+\hat{J}_+}$,  where $\eta_0,\eta_+,\eta_-\,\in\mathbb{C}$ 
and satisfy the following relations. 
\begin{align*}
\vert e^{\eta_0}\vert^2=\frac{1}{1+\vert \eta_+\vert^2},\,\,
 e^{2i\,\text{Im}\eta_0}=-\frac{\bar{\eta}_-}{\eta_+}.
\end{align*}
Noticing that $e^{\eta_+J_{+}}\ket{j,j}=\ket{j,j}$, and quotienting out by the isotropy subgroup 
$H$ corresponds to the choice $\eta_0\in\mathbb{R}$, the corresponding $SU(2)$ group element 
is given by $e^{\zeta\hat{J}_-}e^{-\ln(1+\vert\zeta\vert^2) \hat{J}_z}e^{-\bar\zeta\hat{J}_+}$.

The coherent state manifold is 
$SU(2)/U(1)=S^2$, and an arbitrary CS is given as
\begin{align}
\ket{\zeta}=\frac{1}{(1+\vert\zeta\vert^2)^j}\exp(\zeta\hat{J}_-)\ket{j,j}. 
\label{eqn:coherentexploweringsu2}
\end{align}
The correspondence with the unit sphere can be made explicit by setting 
$\zeta=e^{i\phi}\tan(\frac{\theta}{2}),\,\,\theta\in[0,\pi],
\phi\in[0,2\pi)$, $\zeta$ being the stereographic projection from
the South Pole of the sphere through $\vec{{n}}=(\cos\phi\sin\theta,\sin\phi\sin\theta,\cos\theta)$
to the complex plane. The fiducial state $\ket{j,j}\equiv\ket{0}$ corresponds to the North Pole, 
$\vec{{n}_0}=(0,0,1)\,(\zeta=0)$. We identify the South Pole with all points $\vert\zeta\vert=\infty$.

The various equivalent definitions for the $SU(2)$ CS we shall use are 
\begin{align}
\ket{\vec{n}}\equiv\ket{\theta,\phi}&\equiv\ket{\zeta}=e^{i\phi j}e^{-i\phi \hat{J}_z}e^{-i\theta \hat{J}_y}\ket{0},\nonumber\\
&=\frac{1}{(1+\vert\zeta\vert^2)^j}\exp(\zeta \hat{J}_-)\ket{0},\,\,\zeta=\tan(\frac{\theta}{2})e^{i\phi},\nonumber\\
&= \exp(-i\theta \vec{\hat{J}}\cdot\vec{{m}})\ket{0},\,\,
\vec{m}=\frac{\vec{{n}_0}\times\vec{{n}}}{\vert\vec{{n}_0}\times\vec{{n}}\vert}.\nonumber\\
\vec{m}&=(-\sin\phi,\cos\phi,0),\quad\vec{{n}}\cdot\vec{{n}_0}=\cos(\theta),\nonumber\\
&=\exp\{\xi \hat{J}_+ - \bar\xi \hat{J}_-\}\ket{0},\,\,\xi=-\frac{\theta}{2}e^{-i\phi}.
\end{align}

The generators of $SU(2)$ create closed curves on $S^2$.

Consider CS along a closed path in $\Omega=S^2$,
\[\{\ket{\zeta(\chi)}:\,\ket{\zeta(\chi_0)}=\ket{\zeta(0)}\}.\]
Using Eqn.~\eqref{eqn:horizontalcurvegeometricphase}, the accumulated geometric phase takes the form
($\zeta$ depends implicitly on $\chi$)
\begin{align}
 &\phi_{\text{geom}}=i\int_{0}^{\chi}d\chi'\braket{\zeta(\chi')}{\frac{\text{d}\zeta(\chi')}{\d\chi'}}
 =i\int_0^{\chi_0}
 \frac{j(\bar\zeta \text{d}\zeta-\zeta \text{d}\bar\zeta)}{1+\vert\zeta\vert^2}\nonumber\\
& =-j\int_0^{\chi_0}(1-\cos\theta)\d\phi\,\quad(\text{Using }\zeta=\tan\frac{\theta}{2}e^{i\phi})\nonumber\\
&=-j\oiint\limits_{S(\mathscr{C})}\sin\theta \d\theta\wedge\d\phi=-j\Omega(\mathscr{C}).
\end{align}
where $\Omega(\mathscr{C})$ is the solid angle subtended by the curve $\mathscr{C}$ at the centre of the sphere. Thus, the geometric
phase is equal to the negative of the symplectic area enclosed by the curve.
Since, we are on a compact surface, the requirement, $e^{-ij\Omega}=e^{ij(4\pi-\Omega)}$ is consistent with 
the quantization of $j\text{ in }\frac{\mathbb{N}}{2}$.

Now, we turn to the construction of eigenstates. Note that 
\begin{align*}
	&\exp(-i\theta\vec{\hat{J}}.\vec{{m}})\hat{J_z}\exp(i\theta\vec{\hat{J}}.\vec{{m}})=\vec{\hat{J}}.\vec{{n}}.\\
	&\vec{n}=(\sin\theta\cos\phi,\sin\theta\sin\phi,\cos\theta),\,\vec{m}=(-\sin\phi,\cos\phi,0).
\end{align*}
Hence, it is sufficient to consider orbits generated by $\hat{J}_z$.  
Using Eqn.~\eqref{eqn:averageclosedorbit}
\begin{align}
	\ket{j,m}\equiv	\ket{\bar{\psi}_{\hat{J}_z,m}}= \mathcal{N}_{m}\int_0^{2\pi}\d\phi&\,e^{im\phi} e^{-i\hat{J}_z\phi}\ket{\theta,0}\nonumber\\
	=\mathcal{N}_{m}\int_0^{2\pi}\d\phi&\, e^{i(m-j)\phi}\ket{\theta,\phi}.
	\label{eqn:orbitgeneraljz}
\end{align}
Here, we have considered orbits $\mathcal{O}_{J_z,\left(\sin\theta,\,0,\,\cos\theta\right)}$ which are latitudes at an angle of $\theta$.
However, in order that $\ket{\bar{\psi}_{\hat{J}_z,m}}\neq 0$, Eqn.~\eqref{eqn:Quantizationarbitraryclosedorbit} enforces
\begin{align}
\phi_{\text{geom}}(2\pi)\,+\, 2\pi(m-j\cos\theta)&=2\pi p,\,p\in\mathbb{Z}\nonumber\\
\implies-2\pi j(1-\cos\theta)\,+\,2\pi(m-j\cos\theta)&=2\pi p.\nonumber\\
m=(j+p).\,p\in\mathbb{Z}&
\label{eqn:quantizationsu2arbitraryorbit}
\end{align}
Thus, geometric phase considerations require
the quantization of eigenvalues of $J_z$
For an in phase superposition, using Eqn.~\eqref{eqn:localinphasesuperpositiongeneral} and $\bra{\vec{n}}\hat{\vec J}\ket{\vec{n}}=j\vec{n}$, 
\begin{align*}
 \langle \hat{J_z}\rangle = m\implies j\cos\theta_0=m,
\end{align*}
which selects privileged orbits located at latitudes $\theta_0$. This also implies 
$\vert m\vert\leq j$, and we have $2j+1$ eigenvalues of $\hat{J}_z$ differing from
$j$ by integers.
\begin{figure}[htbp]
    \centering
    \includegraphics[width=0.35\textwidth]{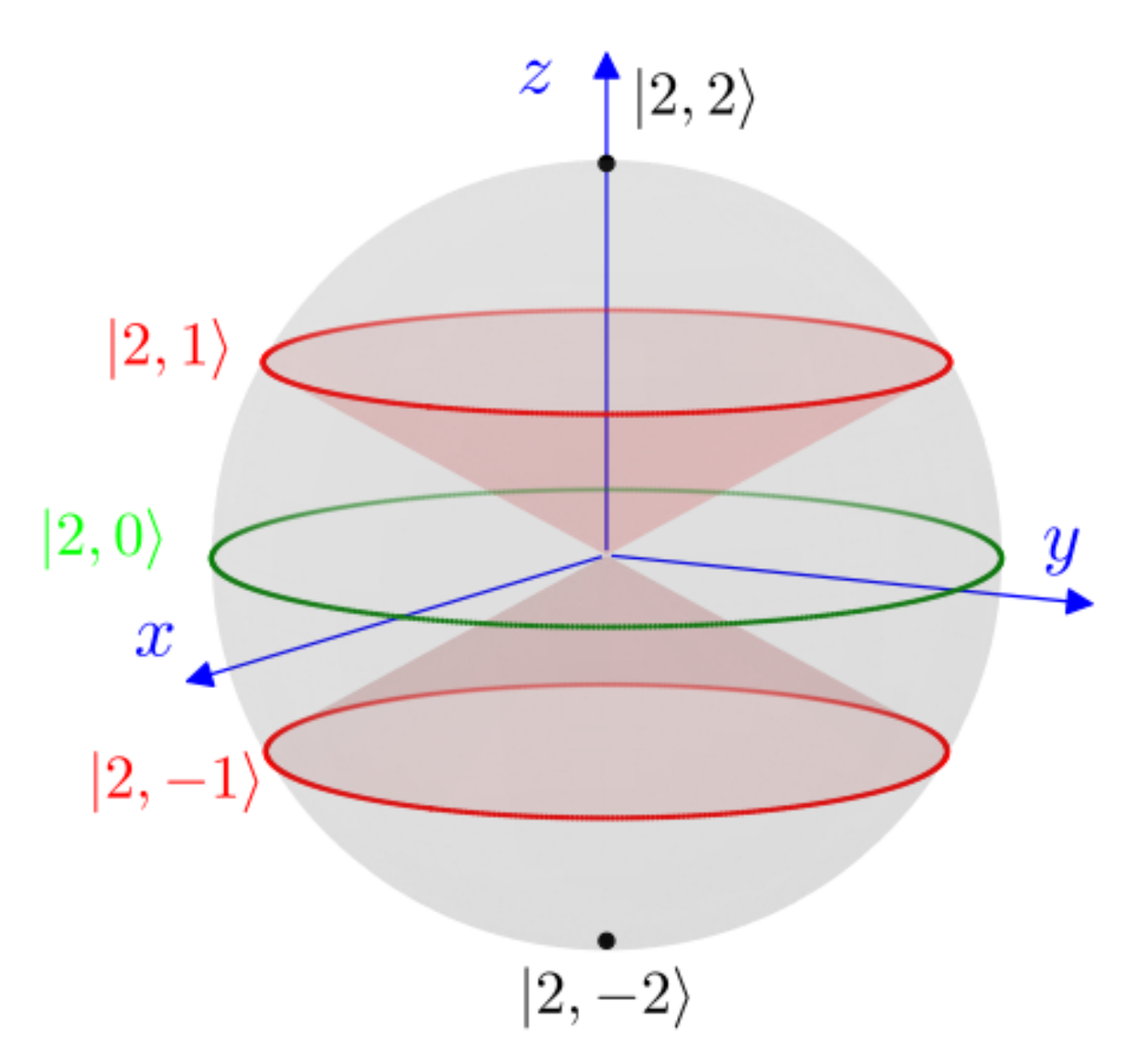}
    \caption{\label{fig:latitudes}
Orbits permitting in phase superposition of eigenstates of $\hat J_z$ for a spin 2 system}
    \end{figure}
In Fig.~\ref{fig:latitudes}, these preferred orbits (latitudes) are shown for the spin 2 representation. 
One can see similar pictures with semiclassical interpretation drawn on spheres with radius $\sqrt{j(j+1)}$.
However, using insights from in phase expansions, Fig.~\ref{fig:latitudes} is exact.

We determine the normalization constant $\mathcal{N}_{m}$ in Eqn.~\eqref{eqn:orbitgeneraljz} by expanding the CS $\ket{\zeta}$.
\[\ket{\zeta}=\frac{1}{\left(1+\vert\zeta\vert^2\right)^j}\sum_{p=0}^{2j}\zeta^p\sqrt{\binom{2j}{p}}\,\,\ket{j,j-p}.\]
 \begin{align}
	 {\mathcal{N}_m}^{-1}=2\pi{\left(\sin\frac{\theta}{2}\right)}^{j-m}{\left(\cos\frac{\theta}{2}\right)}^{j+m}\sqrt{\binom{2j}{j+m}}.  
\end{align}
 We expect that the maximum of the $Q$ distribution lies along
the line of in phase superposition. We verify this for the state $\ket{j,m}$
\begin{align}
	&Q_{\ket{\psi}}(\theta,\phi)=\vert\braket{{\theta,\phi}}{{\psi}}\vert^2,\nonumber\\ 
	&Q_{\ket{j,m}}(\theta,\phi)={\binom{2j}{j+m}}
\left[\cos^{j+m}\frac{\theta}{2}\sin^{j-m}\frac{\theta}{2}\right]^2.\nonumber 
\end{align}
One can easily check that the maximum of $Q_{\ket{j,m}}(\theta,\phi)$ lies at $\theta_0$ satisfying
$j\cos\theta_0=m$, as we anticipated. Equivalently, at $\theta_0$, the superposition coefficients i.e. $\mathcal{N}_m$ is 
minimized.

\subsection{\texorpdfstring{$su(1,1)$}{su(1,1)} coherent states}
The group $SU(1,1)$ consists of the set of complex valued $2\times 2$ unimodular matrices which acting on the complex vector
$(z_1,z_2)^T$ preserves $\vert z_1\vert^2-\vert z_2\vert^2$. A general element can be written as
$\begin{pmatrix}\alpha&\beta\\\bar\beta&\bar\alpha\end{pmatrix}$ with $|\alpha|^2-|\beta|^2=1$. Thus the group
manifold can be described by $x^2+y^2-z^2-w^2=1$, $x,y,z,w\in\mathbb{R}$.

The generators $\hat K_0,\hat K_1,\hat K_2$ of $SU(1,1)$ and their complex combinations $\hat K_{\pm}=K_1\pm i K_2$ obey
\begin{align}
&[\hat K_1,\hat K_2]=-i\hat K_0,~~[\hat K_0,\hat K_{1/2}]=\pm i\hat K_{2/1}, \label{eqn:genereratorssu11}\\
&\hat K_{\pm}=(\hat K_1\pm i\hat K_2),\,[\hat K_0,\hat K_{\pm}]=\pm \hat K_{\pm},\,[\hat K_+,\hat K_-]=-2\hat K_0.\nonumber	
\end{align}
$SU(1,1)$ allows a variety of representations \cite{Bargmann_1947,CoherentPerelomov},
however the ones we consider belong to the so-called positive discrete series.
A representation is described by the Bargmann index $k$ which 
determines the quadratic Casimir $\hat C_2=\hat K_0^2-\hat K_1^2-\hat K_2^2=k(k-1)\mathbbm{1}$. $SU(1,1)$ being non-compact, all unitary representations are infinite dimensional. The basis states $\ket{k,m}$ are eigenkets of $\hat K_0$.
$\hat K_0\ket{k,m}=(k+m)\ket{k,m}$. For the positive discrete series,
$m\in\mathbb{N}\cup\{0\}$.

An arbitrary element of $SU(1,1)$ can be written as $e^{\eta_+\hat K_+}e^{2\eta_0\hat K_0}e^{\eta_-\hat K_-};~~   
\eta_0,\eta_+,\eta_-\in\mathbb{C}$, with the following relations between them
\begin{align*}
	\frac{\eta_+}{\bar{\eta}_-}=-e^{2i\text{Im}(\eta_0)},~~\vert e^{\eta_0}\vert^2=1-\vert\eta_-\vert^2.
\end{align*}
We choose the lowest weight state $\ket{k,0}$ in the representation as the fiducial state.  It has isotropy subalgebra
$\{\mathbbm{1},\hat K_0,\hat K_-\}$. The maximal isotropy subgroup
is $e^{i\eta K_0}$, this coupled with $\hat K_-\ket{k,0}=0$ implies we can confine ourselves to $\eta_0\in\mathbb{R}$.
The CS manifold is thus given by $SU(1,1)/U(1)$.
Introduction to $SU(1,1)$ CS can be found in 
\cite{perelomov1972coherent,CoherentPerelomov,Novaes_2004}.

$SU(1,1)$ CS with fiducial state $\ket{0}\equiv\ket{k,0}$ take the form 
\begin{align}
&\ket{\zeta}=\left(1-|\zeta|^2\right)^k e^{\zeta \hat{K}_+}\ket{0},~~|\zeta|<1,~\zeta\in\mathbb{C}.\label{eqn:su11coherentstatedef1}\\
&\zeta=\tanh\frac{\tau}{2}e^{i\phi},~~
\tau\in[0,\infty),~~\phi\in[0,2\pi).\nonumber
\end{align}
The above parametrization gives the CS as a point on the Poincar\'{e} disk $\zeta$, $|\zeta|<1$.
Alternate parametrizations we will find useful are
\begin{align}
	\ket{\zeta}&=e^{-i\phi k}e^{i\phi \hat K_0}e^{i\tau \hat K_2}\ket{0};\label{eqn:su11eulerangle}\\
	   &=\exp\left(\xi\hat K_{+}-\bar\xi\hat K_-\right)\ket{0},~~\xi=\frac{\tau}{2}e^{i\phi};\label{eqn:su11squeezing}\\
	   &=\exp\left[i\tau\left(\vec{m}\cdot\vec{\hat{K}}\right)\right]\ket{0},\,\vec{m}=\left(\sin\phi,\cos\phi,0\right),\label{eqn:su11axisangle}
\end{align}
where $\vec{\hat{K}}=(\hat{K_1},\hat{K_2},\hat{K_0})$.

$\tau,\phi$ define points on the (upper) hyperboloid $z^2-x^2-y^2=1$, $z>0$ through
\begin{align}
&z=\cosh\tau,~~x=\cos\phi\sinh\tau,~~ y=\sin\phi\sinh\tau.\label{eqn:hyperboloidcoordinates}
\end{align}
The Poincar\'e disk $\zeta$ is the stereographic projection of the upper hyperboloid in the $x-y$ plane through the point $(0,0,-1)$.
We move from the co-ordinate $(x,y,z)$ on the hyperboloid to the point 
$\zeta$ on the Poincar\'e disk by $\zeta=\dfrac{x+iy}{1+z}$.

The geometric phase associated with the closed curve $\mathscr{C}=\{\ket{\zeta(\chi)}:~\ket{\zeta(\chi_0)}=\ket{\zeta(0)}\}$ of CS is
\begin{align}
&\phi_{\text{geom}}= i\int_{0}^{\chi_0}\d\chi\,\bra{\zeta(\chi)}\frac{\text{d}}{\text{d}\chi}\ket{\zeta(\chi)}\nonumber\\
&=\oint\frac{i\,k}{1-\vert\zeta\vert^2}\left[\bar{\zeta}\d\zeta-\zeta{\text{d}\bar\zeta}\right]
	=-k\oiint\limits_{S(\mathscr{C})}\frac{2i\,~\text{d}\zeta\text{d}\bar\zeta}{\left(1-\vert\zeta\vert^2\right)^2}\nonumber\\
	&=-k\oint\left(-1+\cosh\tau\right)\d\phi= -k\oiint\limits_{S(\mathscr{C})}\sinh\tau\d\tau\wedge\d\phi.
\label{eqn:geomphasesu11}
\end{align}
In the last line of Eqn.~\eqref{eqn:geomphasesu11} $\phi$ and $\tau$ are implicit functions of $\chi$. The geometric phase is again given as the negative symplectic area enclosed by $\mathscr{C}$ in the CS manifold.

Among the various useful applications of $SU(1,1)$, one occurs in quantum optics in the context of squeezed states. 
For example a possible realization of the algebra in
terms of a single mode of the electromagnetic field is
\begin{align}
	\hat K_+=\frac{1}{2}\left(\hat{a}^{\dagger}\right)^2,~~\hat K_-=\frac{1}{2}\hat{a}^2,~~\hat K_0=\frac{1}{4}\left(\hat{a}\hat{a}^\dagger+\hat{a}^\dagger\hat{a}\right).
\label{eqn:generatorssu11singlemode}
\end{align}
The quadratic Casimir for this realization evaluates to $-\frac{3}{16}\mathbbm{1}=k(k-1)\mathbbm{1}$. This corresponds to 
the Bargmann index $k=\frac{1}{4}/\frac{3}{4}$ for
the choice of the fiducial state $\ket{0}/\ket{1}$ (the ground state/the first excited Fock state), both of which have the isotropy subalgebra 
$\{\mathbbm{1},\hat{K}_-,\hat{K_0}\}$. Other two mode realizations of this algebra which realize the discrete series are also possible.

Now, as in the previous examples let us turn to the construction of eigenstates. 
We consider three classes of generators: the elliptic ($\hat{K}_0$) , hyperbolic ($\hat{K}_1,\hat{K}_2$) and parabolic 
($\hat{K}_0+\hat{K}_1,\,\hat{K}_0+\hat{K}_2$)
elements.
\begin{figure*}[htbp!]
\subfigure[t][{\label{fig:hyperbolicheatmap}(colour online) $Q$ function of eigenstate of $\hat{K}_2$ with eigenvalue 2 expressed as in phase superposition along its
orbit so that $\langle\hat{K}_2\rangle=2$.
The location of the states are denoted by red dots. $Q$ function maximum lies along the 
superposition. Bargmann index $k$=3. 
}]{ \includegraphics[width=0.40\textwidth]{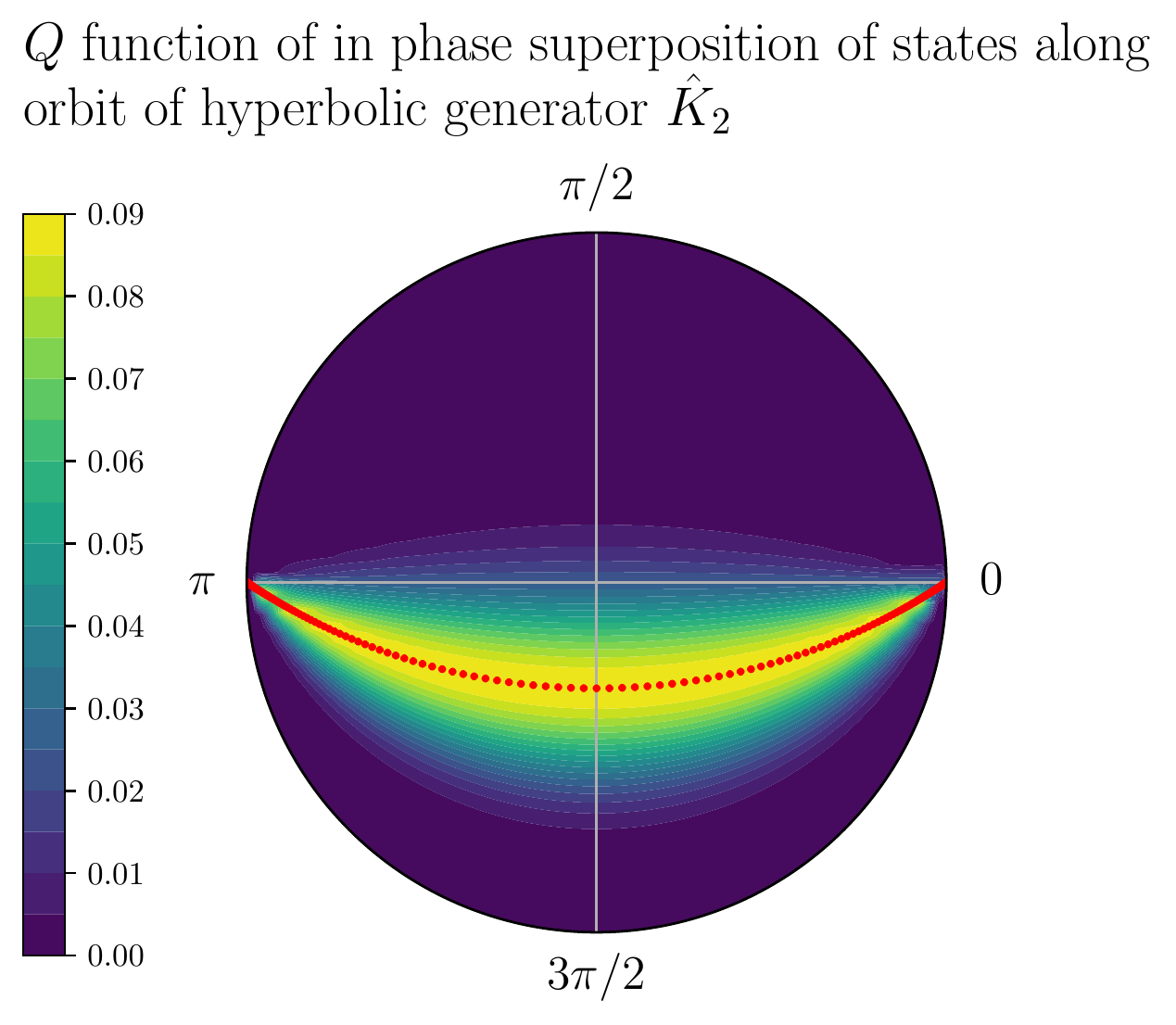}}\hfill
\subfigure[t][{\label{fig:hyperbolicsection}(colour online) Variation of $Q$ function of eigenstate of $\hat{K}_2$ pictured in
Fig.~\ref{fig:hyperbolicheatmap} along the $y$ axis. The $y$ co-ordinate of the $Q$ function maximum (dotted blue line) coincides with the $y$ co-ordinate 
at which the superposition
intersects the $y$ axis (solid black line), $\zeta=i\tanh\frac{\tau_0}{2}$ in Eqn.~\eqref{eqn:inphaseK2condn}, thus verifying our conjecture.
}]{ \includegraphics[width=0.46\textwidth]{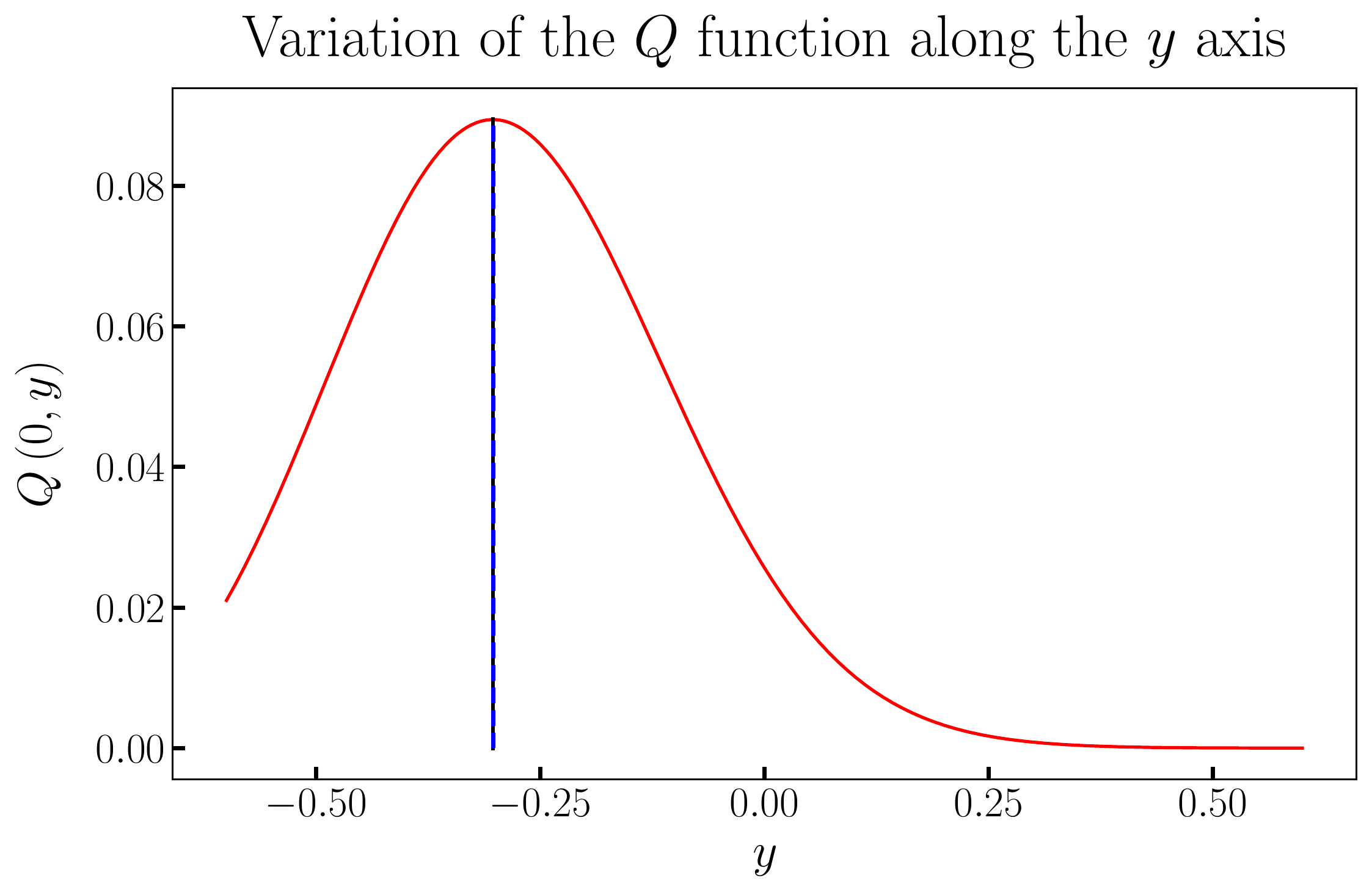}}
\caption{\label{fig:hyperbolic}Heat map of the $Q$ function of eigenstate of $\hat{K}_2$ expressed as in phase superposition along its orbit. }
\end{figure*}

First, let us consider $K_0$ (elliptic element). It generates circles on the Poincar\'{e} 
disk.
\[ e^{i\phi\hat{K_0}}\ket{\zeta}=e^{i\phi k}\ket{\zeta e^{i\phi}}.\]

We construct eigenstates as an expansion over the circle of radius $\tanh\frac{R}{2};~R>0$ on the Poincar\'e disk. In our notation, we are considering the 
orbit $\mathcal{O}_{K_0,\,\tanh\frac{R}{2}}$ (in co-ordinates $\zeta$) which is $\vert\zeta\vert=\tanh\frac{R}{2}$.

Using Eqn.~\eqref{eqn:averageclosedorbit}, the eigenstate with eigenvalue $m_1$ is
\begin{align}
	\ket{\bar{\psi}_{\hat{K_0},m_1}}=\mathcal{N}_{m_1}\int_0^{2\pi}\d\phi\,e^{i(m_1-k)\phi}\,\ket{\tanh\frac{R}{2}~e^{-i\phi}}.
\label{eqn:eigenstateK0intermediate}
\end{align}
Since the Pancharatnam phase must be quantized over an orbit, we get from Eqn.~\eqref{eqn:Quantizationarbitraryclosedorbit} using 
$\langle K_0\rangle=k\cosh R$
\begin{align*}
2\pi k\left(\cosh R -1\right) &+2\pi\left(m_1-k\cosh R\right)\in2\pi\mathbb{Z},\\
\implies m_1-k&=m\in\mathbb{Z}.
\end{align*}
which is the expected result for the quantization of $m$.

From Eqn.~\eqref{eqn:localinphasesuperpositiongeneral}, in phase superposition occurs on special orbits of radius $\text{tanh}\frac{R_0}{2}$ where
$R_0$ satisfies
\begin{align}
&m_1=k\cosh R_0=m+k,
\label{eqn:su11inphasecondition}\\
&\implies m_1\geq k\implies m\in \mathbb{N}\cup\{0\}.\nonumber
\end{align}
Thus, the eigenstate $\ket{k,m}$ of $K_0$ with eigenvalue $k+m$ is expressed as 
\begin{align}
	\ket{k,m}=\mathcal{N}_{m}\int_0^{2\pi}\d\phi\,e^{im\phi}\,\ket{\tanh\frac{R}{2}~e^{-i\phi}},
\label{eqn:eigenstateK0}
\end{align}
where the normalization $\mathcal{N}_m$ can be calculated using 
$\ket{\zeta}=(1-|\zeta|^2)^k\sum\limits_{m=0}^{\infty}\sqrt{\frac{\Gamma(2k+m)}{m!\Gamma(2k)}}\zeta^m\ket{k,m}$.
\begin{align*}
	\mathcal{N}_m^{-1}=2\pi\sqrt{\frac{\Gamma(2k+m)}{m!\Gamma(2k)}}\left(\tanh\frac{R}{2}\right)^m\left(\text{sech}\frac{R}{2}\right)^{2k}.
\end{align*}
The $Q$ function of $\ket{k,m}$  is
\begin{align}
Q_{\ket{k,m}}&=
\vert\braket{k,m}{\zeta}\vert^2\nonumber\\
&=\frac{\Gamma(2k+m)}{m!\Gamma(2k)}\left(\tanh \frac{R}{2}\right)^{2m}\left(\text{sech}\frac{R}{2}\right)^{4k}.
\label{eqn:Qfuncsu11K0}
\end{align}
As expected from the conjecture, $Q_{\ket{k,m}}$ is maximized at $k\cosh R_0=m+k$, which is the circle of in phase superposition in 
Eqns.~\eqref{eqn:eigenstateK0intermediate},\eqref{eqn:su11inphasecondition} whereas $\mathcal{N}_m$ is minimized.

%
%
%
\begin{figure*}[htbp!]
\subfigure[t][{\label{fig:parabolicheatmap}(colour online) $Q$ function of eigenstate of $\hat{K}_0 + \hat{K}_1$ with eigenvalue 2 expressed as in phase superposition along its 
orbit so that $\langle\hat{K}_0+\hat{K}_1\rangle=2$ .
The location of the states are denoted by red dots. $Q$ function maximum lies along the 
superposition. Bargmann index $k$=3. 
}]{ \includegraphics[width=0.42\textwidth]{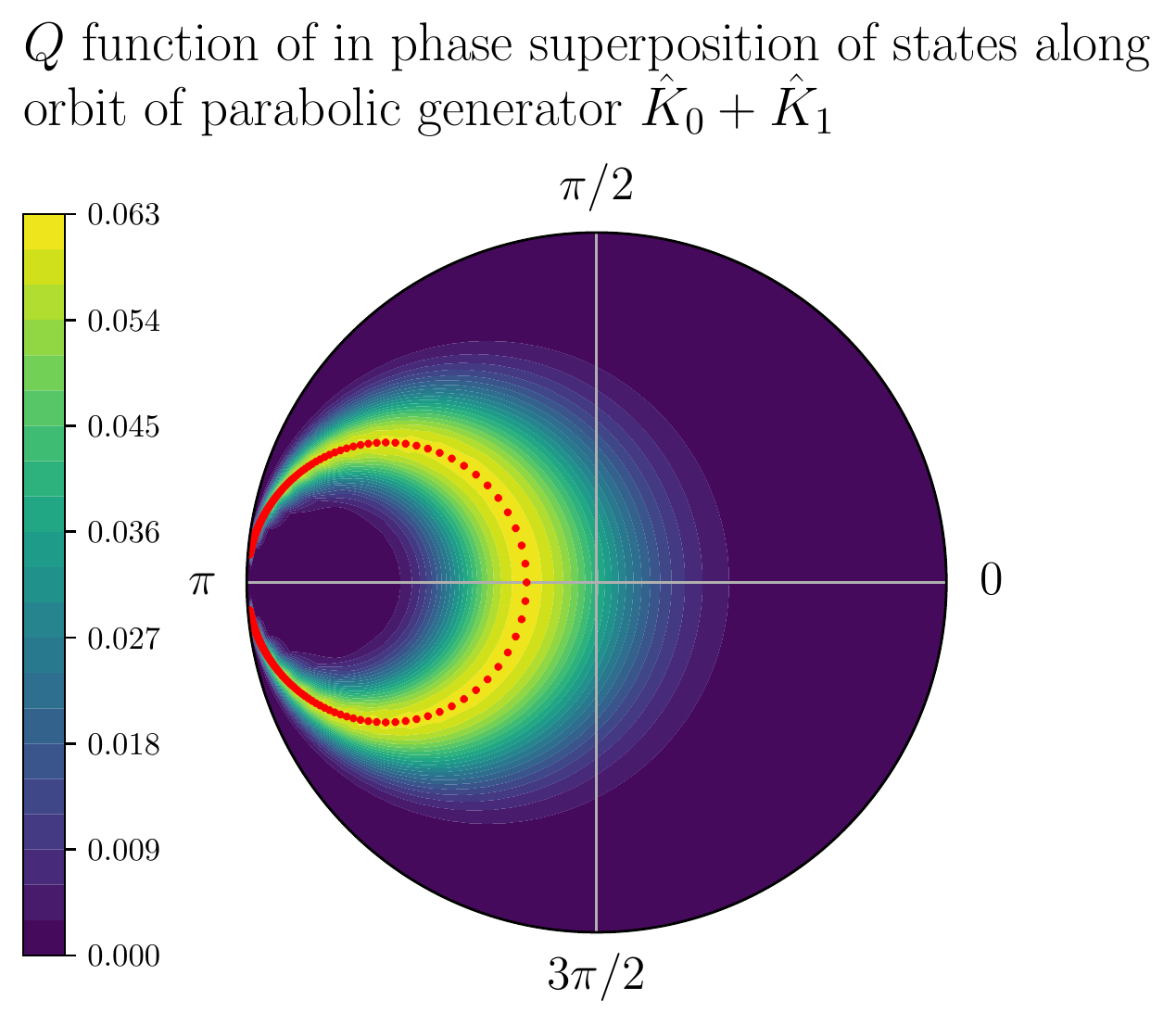}}\hfill
\subfigure[t][{\label{fig:parabolicsection}(colour online) Variation of $Q$ function of eigenstate of $\hat{K}_0+\hat{K}_1$ pictured in
Fig.~\ref{fig:parabolicheatmap} along the $x$ axis. The $x$ co-ordinate of the $Q$ function maximum (dotted blue line) coincides with the $x$ co-ordinate 
at which the superposition
intersects the $x$ axis (solid black line), $\zeta=\tanh\frac{\tau_0}{2}$ in Eqn.~\eqref{eqn:inphasecondnparabolic}, thus verifying our conjecture.
}]{ \includegraphics[width=0.46\textwidth]{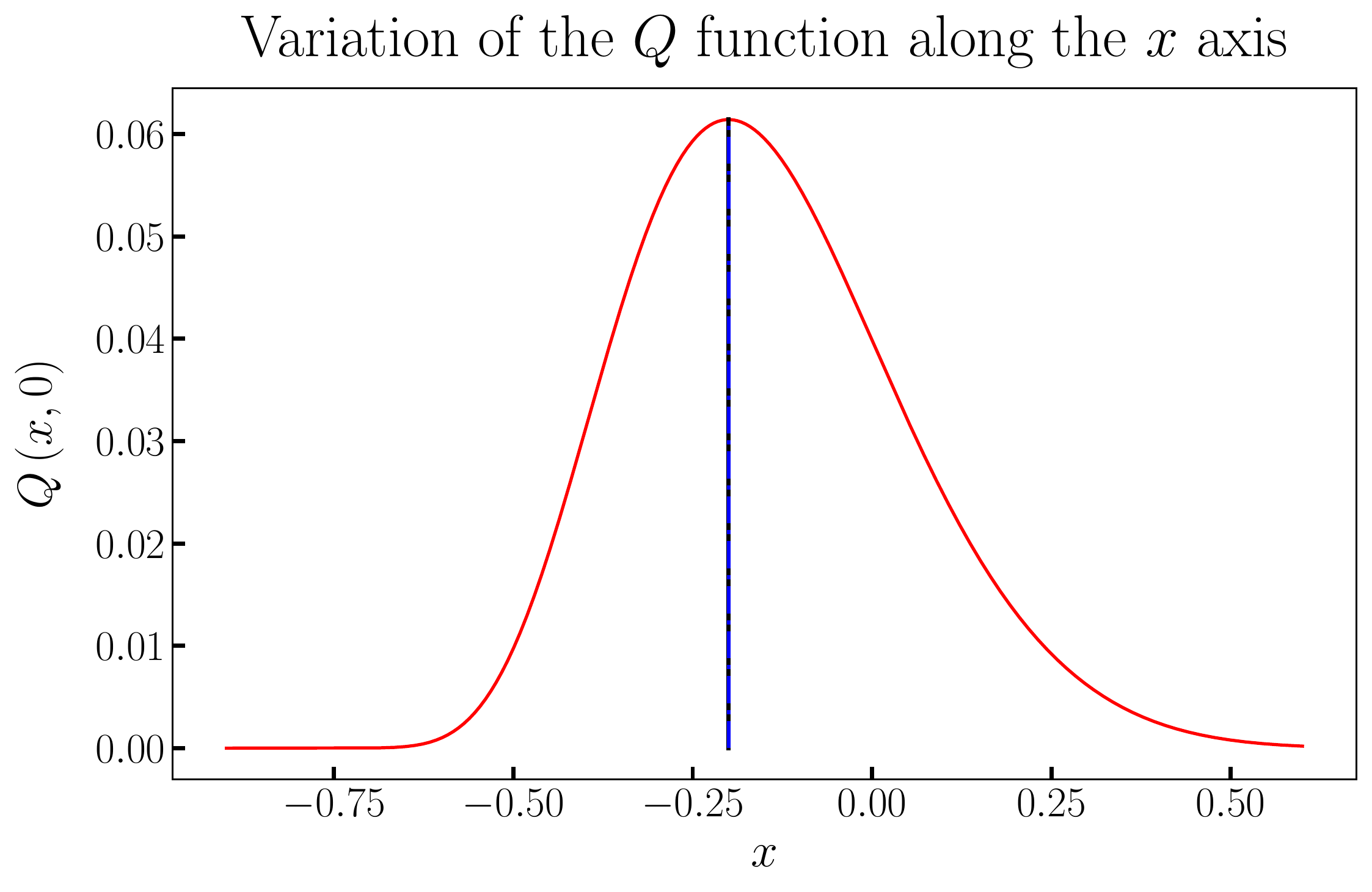}}
\caption{\label{fig:parabolic}Heat map of the $Q$ function of eigenstate of $\hat{K}_0+\hat{K}_1$ expressed as in phase superposition along its orbit. }
\end{figure*}
Let us now construct the eigenstates of the non-compact generators.
Note that $\hat{K_1}$ and $\hat{K_2}$ are related by conjugation,
\[e^{-i\frac{\pi}{2}\hat{K_0}}\hat{K_1}e^{i\frac{\pi}{2}\hat{K_0}}=\hat{K_2}.\]
Hence, for the hyperbolic class we concentrate on the eigenstates generated by $\hat{K_2}$.

Using 
\begin{align}
\label{eqn:squeeze_composition}
&S(\zeta_1)S(\zeta_2)=S(\zeta_3)e^{i\phi\hat{K}_3},\\
&\zeta_3=\frac{\zeta_1+\zeta_2}{1+\bar{\zeta_1}\zeta_2},\,\phi=2\arg(1+\zeta_1\bar{\zeta}_2),\nonumber
\end{align}
where $S(\zeta)=\left( 1-\vert\zeta\vert^2\right)^k e^{\zeta \hat{K}_+}$.

The action of $e^{-i\chi K_2}$ on the CS $\ket{\zeta}$ is
\begin{align*}
&e^{-i\chi\hat{K}_2}\ket{\zeta}=e^{2ik\arg(1-\tanh\frac{\chi}{2}\bar{\zeta})}
\Big\rvert\frac{\zeta-\tanh\frac{\chi}{2}}{1-\zeta\tanh\frac{\chi}{2}}\Big\rangle.
\end{align*}
On the hyperboloid it produces boosts in the $(x-z)$ plane.

The equation of the orbits can be determined by noting that $\langle\hat{K}_2\rangle$ remains constant on them.
\[\langle\hat{K}_2\rangle=\frac{2k\,\text{Im}\,{\bar\zeta}}{1-\vert\zeta\vert^2}=2k\sinh\tau\sin\phi.\]
Hence the orbits of $K_2$ given in terms of $\tau$ and $\phi$ are 
\[\sinh\tau\sin\phi=\sinh\tau_0.\] At $\tau=\tau_0$, $\phi=\frac{\pi}{2}$ the orbit intersects the vertical axis of the Poincar\'e disc
at $\zeta_0=i\tanh\frac{\tau_0}{2}$, and these orbits pass through the diameter at $\pm 1$. (See the red dotted line in Fig.~\ref{fig:hyperbolicheatmap}.)

Thus, applying Eqn.~\eqref{eqn:generalopenorbitaverage} we get the eigenstate of $\hat{K}_2$ with eigenvalue $t_0$ $\ket{\bar{\psi}_{\hat{K}_2,t_0}}$
expressed as a superposition over the orbit $\mathcal{O}_{K_2,i\tanh\frac{\tau_0}{2}}$.
\begin{align}
&\ket{\bar{\psi}_{\hat{K}_2,t_0}} =\mathcal{N}\int\limits_{-\infty}^{\infty}\d\chi\,e^{it_0\chi}e^{i\phi(\chi,\tau_0)}\ket{\zeta(\chi,\tau_0)},\label{eqn:genorbitK2eigenstate}\\
&\phi(\chi,\tau_0)=2k\arg\left(1+i\tanh\frac{\chi}{2}\tanh\frac{\tau_0}{2}\right),\nonumber\\
&\zeta(\chi,\tau_0)=\frac{i\tanh\frac{\tau_0}{2}-\tanh\frac{\chi}{2}}{1-i\tanh\frac{\chi}{2}\tanh\frac{\tau_0}{2}}.\nonumber
\end{align}
We get an in phase superposition when $\langle\hat{K}_2\rangle=t_0$.
This determines $\tau_0$ through
\begin{align}
\sinh\tau_0=-\frac{t_0}{k}.
\label{eqn:inphaseK2condn}
\end{align}

Since $\sinh\tau\in\mathbb{R}$, the spectrum of $\hat{K}_2$ consists of the entire real line. Unlike the other cases, obtaining a closed form expression
for the $Q$ function of the eigenstate of $\hat{K}_2$ is non-trivial. Hence, in this article, to prove the validity of the conjecture
we plot the $Q$ function of a discretized version of the in phase superposition using Eqns.~\eqref{eqn:genorbitK2eigenstate} and ~\eqref{eqn:inphaseK2condn}
in Fig.~\ref{fig:hyperbolic}\,.
We see that the maximum of the $Q$ function does indeed lie along the superposition as posited.

Finally, we turn to the parabolic class, $\hat{K}_0+\hat{K}_1$.

The orbits are determined as the loci of points on which $\langle \hat{K}_0+\hat{K}_1\rangle$ stays constant.
In $\tau-\phi$ co-ordinates, this translates to 
\begin{align}
&\cosh\tau\left(1+\tanh\tau\cos\phi\right)=\exp(\tau_0).
\label{eqn:parabolicorbiteqn}
\end{align}
$\tau_0$ determines the point at which the orbit crosses the x-axis at $\tanh\dfrac{\tau_0}{2}$.
Thus Eqn.~\eqref{eqn:parabolicorbiteqn} specifies the orbit
$\mathcal{O}_{K_0+K_1,\tanh(\tau_0/2)}$.

They are circular orbits that pass through -1 and are tangential to the unit circle at -1. 
(See the red dotted circle in Fig.~\ref{fig:parabolicheatmap}.)
These are also called horocycles.
\begin{align}
&e^{-i\chi(\hat{K}_0+\hat{K}_1)}=S\left[\zeta(\chi)\right]e^{i\phi(\chi)\hat{K}_0},\label{eqn:parabolicsqueezingelement}\\
&\zeta(\chi)=\sqrt{\frac{\chi^2}{4+\chi^2}}\exp\left[i\left(\frac{\phi(\chi)}{2}-\frac{\pi}{2}+\arg(\chi)\right)\right],\nonumber\\
&\phi(\chi)=2\arg\left(1-i\frac{\chi}{2}\right).\nonumber
\end{align}
Finally, using Eqns.~\eqref{eqn:generalopenorbitaverage},\eqref{eqn:squeeze_composition},\eqref{eqn:parabolicsqueezingelement}
we get the eigenstate of $\hat{K}_0+\hat{K}_1$ with eigenvalue $t_0$ expressed as a superposition over the orbit 
$\mathcal{O}_{K_0+K_1,\,\tanh\frac{\tau_0}{2}}$ in Eqn.~\eqref{eqn:parabolicgeneralorbit}.
\begin{widetext}
\begin{equation}
\ket{\psi_{\hat{K}_0+\hat{K}_1,t_0}}=\mathcal{N}\int\limits_{-\infty}^{\infty}\d \chi\, 
\Bigg\rvert\frac{\zeta(\chi)+\tanh\frac{\tau_0}{2}e^{i\phi(\chi)}}
{1+\overline{\zeta(\chi)}\tanh\frac{\tau_0}{2}e^{i\phi(\chi)}}\Bigg\rangle
\exp \left(i\left[t_0\chi+k\phi(\chi)+2k\arg\left(1+\zeta(\chi)\tanh\frac{\tau_0}{2}e^{-i\phi(\chi)}\right)\right]\right).
\label{eqn:parabolicgeneralorbit}
\end{equation}
\end{widetext}
We get in phase superposition when $\langle \hat{K}_0+\hat{K}_1\rangle=t_0 $. This determines $\tau_0$ through
\begin{align}
\exp\left(\tau_0\right)=\frac{t_0}{k}.
\label{eqn:inphasecondnparabolic}
\end{align}
Hence, we know that the spectrum of the parabolic operator consists of the positive real line.

Again, the point of deriving these equations is to show that the $Q$ function maximum of the eigenstate 
lies along the in phase superposition. We find this analytically
intractable. So, we discretize Eqn.~\eqref{eqn:parabolicgeneralorbit} using the condition in Eqn.~\eqref{eqn:inphasecondnparabolic}. The resulting
plot is shown in Fig.~\ref{fig:parabolic}. We see that the maximum of the $Q$ function indeed lies along the superposition, 
in line with our conjecture.
\section{Conclusion}
In this article, we have shed light on how we can use superpositions of coherent states along group orbits to 
construct the eigenstates of the corresponding generator. While this construction applies to any orbit, in phase
orbits exhibit many distinguishing characteristics. 

One sees integral quantization of geometric phase for in phase closed orbits. While this idea has been around in geometric quantization
literature, we bring an alternate and perhaps more approachable perspective to the problem.

We see maximization of $Q$ function along both open/closed in phase orbits in a number of illustrative examples, however we cannot
furnish a proof at this point. We also provide some intuition which guided us to this conjecture using interference in phase space ideas.

It would be good to see a proof/disproof of the conjecture. As already remarked, we do not have an example where this works for a manifold which is
not Hermitian symmetric.

We hope that our article will be of interest both to people interested in quantum state engineering/quantum optics and also more 
mathematically inclined physicisits.

\section{Acknowledgements}
The author thanks Prof.\,R. Simon at the Institute of Mathematical Sciences, Chennai (IMSc) for 
useful suggestions and sharing unpublished work.~\cite{simontalk} A portion of this work was done when the author 
was a visitor at IMSc .
\bibliography{refs}

\begin{thebibliography}{46}%
\makeatletter
\providecommand \@ifxundefined [1]{%
 \@ifx{#1\undefined}
}%
\providecommand \@ifnum [1]{%
 \ifnum #1\expandafter \@firstoftwo
 \else \expandafter \@secondoftwo
 \fi
}%
\providecommand \@ifx [1]{%
 \ifx #1\expandafter \@firstoftwo
 \else \expandafter \@secondoftwo
 \fi
}%
\providecommand \natexlab [1]{#1}%
\providecommand \enquote  [1]{``#1''}%
\providecommand \bibnamefont  [1]{#1}%
\providecommand \bibfnamefont [1]{#1}%
\providecommand \citenamefont [1]{#1}%
\providecommand \href@noop [0]{\@secondoftwo}%
\providecommand \href [0]{\begingroup \@sanitize@url \@href}%
\providecommand \@href[1]{\@@startlink{#1}\@@href}%
\providecommand \@@href[1]{\endgroup#1\@@endlink}%
\providecommand \@sanitize@url [0]{\catcode `\\12\catcode `\$12\catcode
  `\&12\catcode `\#12\catcode `\^12\catcode `\_12\catcode `\%12\relax}%
\providecommand \@@startlink[1]{}%
\providecommand \@@endlink[0]{}%
\providecommand \url  [0]{\begingroup\@sanitize@url \@url }%
\providecommand \@url [1]{\endgroup\@href {#1}{\urlprefix }}%
\providecommand \urlprefix  [0]{URL }%
\providecommand \Eprint [0]{\href }%
\providecommand \doibase [0]{http://dx.doi.org/}%
\providecommand \selectlanguage [0]{\@gobble}%
\providecommand \bibinfo  [0]{\@secondoftwo}%
\providecommand \bibfield  [0]{\@secondoftwo}%
\providecommand \translation [1]{[#1]}%
\providecommand \BibitemOpen [0]{}%
\providecommand \bibitemStop [0]{}%
\providecommand \bibitemNoStop [0]{.\EOS\space}%
\providecommand \EOS [0]{\spacefactor3000\relax}%
\providecommand \BibitemShut  [1]{\csname bibitem#1\endcsname}%
\let\auto@bib@innerbib\@empty
\bibitem [{\citenamefont {Schr{\"o}dinger}(1926)}]{schrodinger1926stetige}%
  \BibitemOpen
  \bibfield  {author} {\bibinfo {author} {\bibfnamefont {E.}~\bibnamefont
  {Schr{\"o}dinger}},\ }\href {\doibase 10.1007/BF01507634} {\bibfield
  {journal} {\bibinfo  {journal} {Naturwissenschaften}\ }\textbf {\bibinfo
  {volume} {14}},\ \bibinfo {pages} {664} (\bibinfo {year} {1926})}\BibitemShut
  {NoStop}%
\bibitem [{\citenamefont {Perelomov}(1972)}]{perelomov1972coherent}%
  \BibitemOpen
  \bibfield  {author} {\bibinfo {author} {\bibfnamefont {A.}~\bibnamefont
  {Perelomov}},\ }\href {\doibase 10.1007/bf01645091} {\bibfield  {journal}
  {\bibinfo  {journal} {Communications in Mathematical Physics}\ }\textbf
  {\bibinfo {volume} {26}},\ \bibinfo {pages} {222} (\bibinfo {year}
  {1972})}\BibitemShut {NoStop}%
\bibitem [{\citenamefont {Berry}(1984)}]{Berry}%
  \BibitemOpen
  \bibfield  {author} {\bibinfo {author} {\bibfnamefont {M.~V.}\ \bibnamefont
  {Berry}},\ }\href {\doibase 10.1098/rspa.1984.0023} {\bibfield  {journal}
  {\bibinfo  {journal} {Proceedings of the Royal Society A: Mathematical,
  Physical and Engineering Sciences}\ }\textbf {\bibinfo {volume} {392}},\
  \bibinfo {pages} {45} (\bibinfo {year} {1984})}\BibitemShut {NoStop}%
\bibitem [{\citenamefont {Pancharatnam}(1956)}]{Pancharatnam}%
  \BibitemOpen
  \bibfield  {author} {\bibinfo {author} {\bibfnamefont {S.}~\bibnamefont
  {Pancharatnam}},\ }\href {\doibase 10.1007/BF03046050} {\bibfield  {journal}
  {\bibinfo  {journal} {Proceedings of the Indian Academy of Sciences - Section
  A}\ }\textbf {\bibinfo {volume} {{44}}},\ \bibinfo {pages} {247–262}
  (\bibinfo {year} {1956})}\BibitemShut {NoStop}%
\bibitem [{\citenamefont {Mukunda}\ and\ \citenamefont
  {Simon}(1993{\natexlab{a}})}]{kinematicformulation1}%
  \BibitemOpen
  \bibfield  {author} {\bibinfo {author} {\bibfnamefont {N.}~\bibnamefont
  {Mukunda}}\ and\ \bibinfo {author} {\bibfnamefont {R.}~\bibnamefont
  {Simon}},\ }\href {\doibase 10.1006/aphy.1993.1093} {\bibfield  {journal}
  {\bibinfo  {journal} {Annals of Physics}\ }\textbf {\bibinfo {volume}
  {228}},\ \bibinfo {pages} {205} (\bibinfo {year}
  {1993}{\natexlab{a}})}\BibitemShut {NoStop}%
\bibitem [{\citenamefont {Mukunda}\ and\ \citenamefont
  {Simon}(1993{\natexlab{b}})}]{kinematicformulation2}%
  \BibitemOpen
  \bibfield  {author} {\bibinfo {author} {\bibfnamefont {N.}~\bibnamefont
  {Mukunda}}\ and\ \bibinfo {author} {\bibfnamefont {R.}~\bibnamefont
  {Simon}},\ }\href {\doibase 10.1006/aphy.1993.1094} {\bibfield  {journal}
  {\bibinfo  {journal} {Annals of Physics}\ }\textbf {\bibinfo {volume}
  {{228}}},\ \bibinfo {pages} {269} (\bibinfo {year}
  {1993}{\natexlab{b}})}\BibitemShut {NoStop}%
\bibitem [{\citenamefont {{Husimi}}(1940)}]{HusimiQ}%
  \BibitemOpen
  \bibfield  {author} {\bibinfo {author} {\bibfnamefont {K.}~\bibnamefont
  {{Husimi}}},\ }\href {\doibase 10.11429/ppmsj1919.22.4_264} {\bibfield
  {journal} {\bibinfo  {journal} {Proceedings of the Physico-Mathematical
  Society of Japan. 3rd Series}\ }\textbf {\bibinfo {volume} {22}},\ \bibinfo
  {pages} {264} (\bibinfo {year} {1940})}\BibitemShut {NoStop}%
\bibitem [{\citenamefont {Wehrl}(1979)}]{wehrl1979relation}%
  \BibitemOpen
  \bibfield  {author} {\bibinfo {author} {\bibfnamefont {A.}~\bibnamefont
  {Wehrl}},\ }\href {\doibase 10.1016/0034-4877(79)90070-3} {\bibfield
  {journal} {\bibinfo  {journal} {Reports on Mathematical Physics}\ }\textbf
  {\bibinfo {volume} {16}},\ \bibinfo {pages} {353} (\bibinfo {year}
  {1979})}\BibitemShut {NoStop}%
\bibitem [{\citenamefont {Gnutzmann}\ and\ \citenamefont
  {\.{Z}yczkowski}(2001)}]{gnutzmann2001renyi}%
  \BibitemOpen
  \bibfield  {author} {\bibinfo {author} {\bibfnamefont {S.}~\bibnamefont
  {Gnutzmann}}\ and\ \bibinfo {author} {\bibfnamefont {K.}~\bibnamefont
  {\.{Z}yczkowski}},\ }\href {\doibase 10.1088/0305-4470/34/47/317} {\bibfield
  {journal} {\bibinfo  {journal} {Journal of Physics A: Mathematical and
  General}\ }\textbf {\bibinfo {volume} {34}},\ \bibinfo {pages} {10123}
  (\bibinfo {year} {2001})}\BibitemShut {NoStop}%
\bibitem [{\citenamefont {Lieb}(1978)}]{lieb1978proof}%
  \BibitemOpen
  \bibfield  {author} {\bibinfo {author} {\bibfnamefont {E.}~\bibnamefont
  {Lieb}},\ }\href {\doibase 10.1007/BF01940328} {\bibfield  {journal}
  {\bibinfo  {journal} {Commun. Math. Phys.}\ }\textbf {\bibinfo {volume}
  {62}},\ \bibinfo {pages} {35} (\bibinfo {year} {1978})}\BibitemShut {NoStop}%
\bibitem [{\citenamefont {Carlen}(1991)}]{CARLEN1991231}%
  \BibitemOpen
  \bibfield  {author} {\bibinfo {author} {\bibfnamefont {E.~A.}\ \bibnamefont
  {Carlen}},\ }\href {\doibase 10.1016/0022-1236(91)90022-W} {\bibfield
  {journal} {\bibinfo  {journal} {Journal of Functional Analysis}\ }\textbf
  {\bibinfo {volume} {97}},\ \bibinfo {pages} {231 } (\bibinfo {year}
  {1991})}\BibitemShut {NoStop}%
\bibitem [{\citenamefont {Lieb}\ and\ \citenamefont
  {Solovej}(2014)}]{lieb2014proof}%
  \BibitemOpen
  \bibfield  {author} {\bibinfo {author} {\bibfnamefont {E.~H.}\ \bibnamefont
  {Lieb}}\ and\ \bibinfo {author} {\bibfnamefont {J.~P.}\ \bibnamefont
  {Solovej}},\ }\href {\doibase 10.1007/s11511-014-0113-6} {\bibfield
  {journal} {\bibinfo  {journal} {Acta Mathematica}\ }\textbf {\bibinfo
  {volume} {212}},\ \bibinfo {pages} {379} (\bibinfo {year}
  {2014})}\BibitemShut {NoStop}%
\bibitem [{\citenamefont {Simon}(2001)}]{simontalk}%
  \BibitemOpen
  \bibfield  {author} {\bibinfo {author} {\bibfnamefont {R.}~\bibnamefont
  {Simon}},\ }\href@noop {} {\enquote {\bibinfo {title} {Interference in phase
  space, geometric phase, and asymptotic expressions for classical
  polynomials},}\ } (\bibinfo {year} {2001}),\ \bibinfo {note} {\text{Talk} at
  CTS, Indian Institute of Science, Bangalore - \textit{Geometric Phases in
  Physics and Foundations of Quantum Mechanics}}\BibitemShut {NoStop}%
\bibitem [{\citenamefont {Khan}\ \emph {et~al.}(2018)\citenamefont {Khan},
  \citenamefont {Chaturvedi}, \citenamefont {Mukunda},\ and\ \citenamefont
  {Simon}}]{khan2018geometric}%
  \BibitemOpen
  \bibfield  {author} {\bibinfo {author} {\bibfnamefont {M.~N.}\ \bibnamefont
  {Khan}}, \bibinfo {author} {\bibfnamefont {S.}~\bibnamefont {Chaturvedi}},
  \bibinfo {author} {\bibfnamefont {N.}~\bibnamefont {Mukunda}}, \ and\
  \bibinfo {author} {\bibfnamefont {R.}~\bibnamefont {Simon}},\ }\href@noop {}
  {\bibfield  {journal} {\bibinfo  {journal} {arXiv e-prints}\ } (\bibinfo
  {year} {2018})},\ \Eprint {http://arxiv.org/abs/1812.07443} {arXiv:1812.07443
  [quant-ph]} \BibitemShut {NoStop}%
\bibitem [{\citenamefont {Wheeler}(1985)}]{Wheeler_1985}%
  \BibitemOpen
  \bibfield  {author} {\bibinfo {author} {\bibfnamefont {J.~A.}\ \bibnamefont
  {Wheeler}},\ }\href@noop {} {\bibfield  {journal} {\bibinfo  {journal}
  {Letters in Mathematical Physics}\ }\textbf {\bibinfo {volume}
  {\href{http://dx.doi.org/10.1007/bf00398159}{10}}},\ \bibinfo {pages} {201}
  (\bibinfo {year} {1985})}\BibitemShut {NoStop}%
\bibitem [{\citenamefont {Schleich}\ and\ \citenamefont
  {Wheeler}(1987)}]{Schleich_1987}%
  \BibitemOpen
  \bibfield  {author} {\bibinfo {author} {\bibfnamefont {W.}~\bibnamefont
  {Schleich}}\ and\ \bibinfo {author} {\bibfnamefont {J.~A.}\ \bibnamefont
  {Wheeler}},\ }\href@noop {} {\bibfield  {journal} {\bibinfo  {journal}
  {Nature}\ }\textbf {\bibinfo {volume}
  {\href{http://dx.doi.org/10.1038/326574a0}{326}}},\ \bibinfo {pages} {574}
  (\bibinfo {year} {1987})}\BibitemShut {NoStop}%
\bibitem [{\citenamefont {Perelomov}(1986)}]{CoherentPerelomov}%
  \BibitemOpen
  \bibfield  {author} {\bibinfo {author} {\bibfnamefont {A.}~\bibnamefont
  {Perelomov}},\ }\href {\doibase 10.1007/978-3-642-61629-7} {\emph {\bibinfo
  {title} {Generalized Coherent States and Their Applications}}}\ (\bibinfo
  {publisher} {Springer Science \& Business Media},\ \bibinfo {year}
  {1986})\BibitemShut {NoStop}%
\bibitem [{\citenamefont {Gilmore}(1974)}]{gilmore1974properties}%
  \BibitemOpen
  \bibfield  {author} {\bibinfo {author} {\bibfnamefont {R.}~\bibnamefont
  {Gilmore}},\ }\href@noop {} {\bibfield  {journal} {\bibinfo  {journal}
  {\href{https://rmf.smf.mx/ojs/rmf/article/view/1046/832}{Revista Mexicana de
  Fisica}}\ }\textbf {\bibinfo {volume} {23}},\ \bibinfo {pages} {143}
  (\bibinfo {year} {1974})}\BibitemShut {NoStop}%
\bibitem [{\citenamefont {Delbourgo}\ and\ \citenamefont
  {Fox}(1977)}]{delbourgo1977maximum}%
  \BibitemOpen
  \bibfield  {author} {\bibinfo {author} {\bibfnamefont {R.}~\bibnamefont
  {Delbourgo}}\ and\ \bibinfo {author} {\bibfnamefont {J.}~\bibnamefont
  {Fox}},\ }\href {\doibase 10.1088/0305-4470/10/12/004} {\bibfield  {journal}
  {\bibinfo  {journal} {Journal of Physics A: Mathematical and General}\
  }\textbf {\bibinfo {volume} {10}},\ \bibinfo {pages} {L233} (\bibinfo {year}
  {1977})}\BibitemShut {NoStop}%
\bibitem [{\citenamefont {Zhang}\ \emph {et~al.}(1990)\citenamefont {Zhang},
  \citenamefont {Feng},\ and\ \citenamefont {Gilmore}}]{GilmoreCoherentReview}%
  \BibitemOpen
  \bibfield  {author} {\bibinfo {author} {\bibfnamefont {W.-M.}\ \bibnamefont
  {Zhang}}, \bibinfo {author} {\bibfnamefont {D.~H.}\ \bibnamefont {Feng}}, \
  and\ \bibinfo {author} {\bibfnamefont {R.}~\bibnamefont {Gilmore}},\ }\href
  {\doibase 10.1103/RevModPhys.62.867} {\bibfield  {journal} {\bibinfo
  {journal} {Rev. Mod. Phys.}\ }\textbf {\bibinfo {volume} {62}},\ \bibinfo
  {pages} {867} (\bibinfo {year} {1990})}\BibitemShut {NoStop}%
\bibitem [{\citenamefont {Onofri}(1975)}]{Onofri_1975}%
  \BibitemOpen
  \bibfield  {author} {\bibinfo {author} {\bibfnamefont {E.}~\bibnamefont
  {Onofri}},\ }\href {\doibase 10.1063/1.522663} {\bibfield  {journal}
  {\bibinfo  {journal} {Journal of Mathematical Physics}\ }\textbf {\bibinfo
  {volume} {16}},\ \bibinfo {pages} {1087} (\bibinfo {year}
  {1975})}\BibitemShut {NoStop}%
\bibitem [{\citenamefont {Aharonov}\ and\ \citenamefont
  {Anandan}(1987)}]{adiabaticity}%
  \BibitemOpen
  \bibfield  {author} {\bibinfo {author} {\bibfnamefont {Y.}~\bibnamefont
  {Aharonov}}\ and\ \bibinfo {author} {\bibfnamefont {J.}~\bibnamefont
  {Anandan}},\ }\href@noop {} {\bibfield  {journal} {\bibinfo  {journal} {Phys.
  Rev. Lett.}\ }\textbf {\bibinfo {volume}
  {\href{http://dx.doi.org/10.1103/physrevlett.58.1593}{58}}},\ \bibinfo
  {pages} {1593} (\bibinfo {year} {1987})}\BibitemShut {NoStop}%
\bibitem [{\citenamefont {Samuel}\ and\ \citenamefont
  {Bhandari}(1988)}]{cyclicevolution}%
  \BibitemOpen
  \bibfield  {author} {\bibinfo {author} {\bibfnamefont {J.}~\bibnamefont
  {Samuel}}\ and\ \bibinfo {author} {\bibfnamefont {R.}~\bibnamefont
  {Bhandari}},\ }\href {\doibase 10.1103/PhysRevLett.60.2339} {\bibfield
  {journal} {\bibinfo  {journal} {Phys. Rev. Lett.}\ }\textbf {\bibinfo
  {volume} {{60}}},\ \bibinfo {pages} {2339} (\bibinfo {year}
  {1988})}\BibitemShut {NoStop}%
\bibitem [{\citenamefont {Berry}(1987)}]{BerryPancharatnam}%
  \BibitemOpen
  \bibfield  {author} {\bibinfo {author} {\bibfnamefont {M.}~\bibnamefont
  {Berry}},\ }\href {\doibase 10.1080/09500348714551321} {\bibfield  {journal}
  {\bibinfo  {journal} {Journal of Modern Optics}\ }\textbf {\bibinfo {volume}
  {34}},\ \bibinfo {pages} {1401} (\bibinfo {year} {1987})}\BibitemShut
  {NoStop}%
\bibitem [{\citenamefont {Bargmann}(1964)}]{Bargmanninvariant}%
  \BibitemOpen
  \bibfield  {author} {\bibinfo {author} {\bibfnamefont {V.}~\bibnamefont
  {Bargmann}},\ }\href {\doibase 10.1063/1.1704188} {\bibfield  {journal}
  {\bibinfo  {journal} {Journal of Mathematical Physics}\ }\textbf {\bibinfo
  {volume} {5}},\ \bibinfo {pages} {862} (\bibinfo {year} {1964})}\BibitemShut
  {NoStop}%
\bibitem [{\citenamefont {Boya}\ \emph {et~al.}(2001)\citenamefont {Boya},
  \citenamefont {Perelomov},\ and\ \citenamefont {Santander}}]{boya2001berry}%
  \BibitemOpen
  \bibfield  {author} {\bibinfo {author} {\bibfnamefont {L.~J.}\ \bibnamefont
  {Boya}}, \bibinfo {author} {\bibfnamefont {A.~M.}\ \bibnamefont {Perelomov}},
  \ and\ \bibinfo {author} {\bibfnamefont {M.}~\bibnamefont {Santander}},\
  }\href {\doibase 10.1063/1.1396837} {\bibfield  {journal} {\bibinfo
  {journal} {Journal of Mathematical Physics}\ }\textbf {\bibinfo {volume}
  {42}},\ \bibinfo {pages} {5130} (\bibinfo {year} {2001})}\BibitemShut
  {NoStop}%
\bibitem [{\citenamefont {Domic}\ and\ \citenamefont
  {Toledo}(1987)}]{domictoledo1987}%
  \BibitemOpen
  \bibfield  {author} {\bibinfo {author} {\bibfnamefont {A.}~\bibnamefont
  {Domic}}\ and\ \bibinfo {author} {\bibfnamefont {D.}~\bibnamefont {Toledo}},\
  }\href {\doibase 10.1007/BF01450839} {\bibfield  {journal} {\bibinfo
  {journal} {Mathematische Annalen}\ }\textbf {\bibinfo {volume} {276}},\
  \bibinfo {pages} {425} (\bibinfo {year} {1987})}\BibitemShut {NoStop}%
\bibitem [{\citenamefont {Clerc}\ and\ \citenamefont
  {{\O}rsted}(2003)}]{clerc2003corrigendum}%
  \BibitemOpen
  \bibfield  {author} {\bibinfo {author} {\bibfnamefont {J.-L.}\ \bibnamefont
  {Clerc}}\ and\ \bibinfo {author} {\bibfnamefont {B.}~\bibnamefont
  {{\O}rsted}},\ }\href {https://projecteuclid.org/euclid.ajm/1098300997}
  {\bibfield  {journal} {\bibinfo  {journal} {Asian J. Math}\ }\textbf
  {\bibinfo {volume} {7}},\ \bibinfo {pages} {269} (\bibinfo {year}
  {2003})}\BibitemShut {NoStop}%
\bibitem [{\citenamefont {Berceanu}(1999)}]{berceanu1999coherent}%
  \BibitemOpen
  \bibfield  {author} {\bibinfo {author} {\bibfnamefont {S.}~\bibnamefont
  {Berceanu}},\ }\href {https://arxiv.org/abs/math/9903190} {\bibfield
  {journal} {\bibinfo  {journal} {arXiv preprint math/9903190}\ } (\bibinfo
  {year} {1999})}\BibitemShut {NoStop}%
\bibitem [{\citenamefont {Berceanu}(2004)}]{berceanu2004geometrical}%
  \BibitemOpen
  \bibfield  {author} {\bibinfo {author} {\bibfnamefont {S.}~\bibnamefont
  {Berceanu}},\ }\href@noop {} {\  (\bibinfo {year} {2004})},\ \Eprint
  {http://arxiv.org/abs/math/0408233} {arXiv:math/0408233 [math.DG]}
  \BibitemShut {NoStop}%
\bibitem [{\citenamefont {Bech}(2017)}]{bech2017canonical}%
  \BibitemOpen
  \bibfield  {author} {\bibinfo {author} {\bibfnamefont {M.~A.}\ \bibnamefont
  {Bech}},\ }\emph {\bibinfo {title} {Canonical kernels on Hermitian symmetric
  spaces}},\ \href
  {https://data.math.au.dk/publications/phd/2017/math-phd-2017-mab.pdf} {Ph.D.
  thesis},\ \bibinfo  {school} {Aarhus University, Department of Mathematics}
  (\bibinfo {year} {2017})\BibitemShut {NoStop}%
\bibitem [{\citenamefont {Rabei}\ \emph {et~al.}(1999)\citenamefont {Rabei},
  \citenamefont {Arvind}, \citenamefont {Mukunda},\ and\ \citenamefont
  {Simon}}]{Rabei_1999}%
  \BibitemOpen
  \bibfield  {author} {\bibinfo {author} {\bibfnamefont {E.~M.}\ \bibnamefont
  {Rabei}}, \bibinfo {author} {\bibnamefont {Arvind}}, \bibinfo {author}
  {\bibfnamefont {N.}~\bibnamefont {Mukunda}}, \ and\ \bibinfo {author}
  {\bibfnamefont {R.}~\bibnamefont {Simon}},\ }\href {\doibase
  10.1103/physreva.60.3397} {\bibfield  {journal} {\bibinfo  {journal}
  {Physical Review A}\ }\textbf {\bibinfo {volume} {60}},\ \bibinfo {pages}
  {3397} (\bibinfo {year} {1999})}\BibitemShut {NoStop}%
\bibitem [{\citenamefont {Mukunda}\ \emph {et~al.}(2003)\citenamefont
  {Mukunda}, \citenamefont {Arvind}, \citenamefont {Ercolessi}, \citenamefont
  {Marmo}, \citenamefont {Morandi},\ and\ \citenamefont {Simon}}]{mukunda2003}%
  \BibitemOpen
  \bibfield  {author} {\bibinfo {author} {\bibfnamefont {N.}~\bibnamefont
  {Mukunda}}, \bibinfo {author} {\bibnamefont {Arvind}}, \bibinfo {author}
  {\bibfnamefont {E.}~\bibnamefont {Ercolessi}}, \bibinfo {author}
  {\bibfnamefont {G.}~\bibnamefont {Marmo}}, \bibinfo {author} {\bibfnamefont
  {G.}~\bibnamefont {Morandi}}, \ and\ \bibinfo {author} {\bibfnamefont
  {R.}~\bibnamefont {Simon}},\ }\href {\doibase 10.1103/PhysRevA.67.042114}
  {\bibfield  {journal} {\bibinfo  {journal} {Phys. Rev. A}\ }\textbf {\bibinfo
  {volume} {67}},\ \bibinfo {pages} {042114} (\bibinfo {year}
  {2003})}\BibitemShut {NoStop}%
\bibitem [{\citenamefont {Bargmann}(1961)}]{bargmann1961hilbert}%
  \BibitemOpen
  \bibfield  {author} {\bibinfo {author} {\bibfnamefont {V.}~\bibnamefont
  {Bargmann}},\ }\href {\doibase 10.1002/cpa.3160140303} {\bibfield  {journal}
  {\bibinfo  {journal} {Communications on Pure and Applied Mathematics}\
  }\textbf {\bibinfo {volume} {14}},\ \bibinfo {pages} {187} (\bibinfo {year}
  {1961})}\BibitemShut {NoStop}%
\bibitem [{\citenamefont {Bargmann}\ \emph {et~al.}(1971)\citenamefont
  {Bargmann}, \citenamefont {Butera}, \citenamefont {Girardello},\ and\
  \citenamefont {Klauder}}]{bargmann1971completeness}%
  \BibitemOpen
  \bibfield  {author} {\bibinfo {author} {\bibfnamefont {V.}~\bibnamefont
  {Bargmann}}, \bibinfo {author} {\bibfnamefont {P.}~\bibnamefont {Butera}},
  \bibinfo {author} {\bibfnamefont {L.}~\bibnamefont {Girardello}}, \ and\
  \bibinfo {author} {\bibfnamefont {J.~R.}\ \bibnamefont {Klauder}},\ }\href
  {\doibase 10.1016/0034-4877(71)90006-1} {\bibfield  {journal} {\bibinfo
  {journal} {Reports on Mathematical Physics}\ }\textbf {\bibinfo {volume}
  {2}},\ \bibinfo {pages} {221} (\bibinfo {year} {1971})}\BibitemShut {NoStop}%
\bibitem [{\citenamefont {{\'{S}}niatycki}(1980)}]{Sniatycki_1980}%
  \BibitemOpen
  \bibfield  {author} {\bibinfo {author} {\bibfnamefont {J.}~\bibnamefont
  {{\'{S}}niatycki}},\ }\href {\doibase 10.1007/978-1-4612-6066-0} {\emph
  {\bibinfo {title} {Geometric Quantization and Quantum Mechanics}}}\ (\bibinfo
   {publisher} {Springer New York},\ \bibinfo {year} {1980})\BibitemShut
  {NoStop}%
\bibitem [{\citenamefont {Kirillov}(1990)}]{Kirillov1990}%
  \BibitemOpen
  \bibfield  {author} {\bibinfo {author} {\bibfnamefont {A.~A.}\ \bibnamefont
  {Kirillov}},\ }\enquote {\bibinfo {title} {Geometric quantization},}\ in\
  \href {\doibase 10.1007/978-3-662-06793-2_2} {\emph {\bibinfo {booktitle}
  {Dynamical Systems IV: Symplectic Geometry and its Applications}}},\ \bibinfo
  {editor} {edited by\ \bibinfo {editor} {\bibfnamefont {V.~I.}\ \bibnamefont
  {Arnol'd}}\ and\ \bibinfo {editor} {\bibfnamefont {S.~P.}\ \bibnamefont
  {Novikov}}}\ (\bibinfo  {publisher} {Springer Berlin Heidelberg},\ \bibinfo
  {address} {Berlin, Heidelberg},\ \bibinfo {year} {1990})\ pp.\ \bibinfo
  {pages} {137--172}\BibitemShut {NoStop}%
\bibitem [{\citenamefont {Cushman}\ and\ \citenamefont
  {{\'{S}}niatycki}(2013)}]{Cushman_2013}%
  \BibitemOpen
  \bibfield  {author} {\bibinfo {author} {\bibfnamefont {R.}~\bibnamefont
  {Cushman}}\ and\ \bibinfo {author} {\bibfnamefont {J.}~\bibnamefont
  {{\'{S}}niatycki}},\ }\href {\doibase 10.1007/s11784-013-0118-3} {\bibfield
  {journal} {\bibinfo  {journal} {Journal of Fixed Point Theory and
  Applications}\ }\textbf {\bibinfo {volume} {13}},\ \bibinfo {pages} {3}
  (\bibinfo {year} {2013})}\BibitemShut {NoStop}%
\bibitem [{\citenamefont {Janszky}\ and\ \citenamefont
  {Vinogradov}(1990)}]{Janzky1}%
  \BibitemOpen
  \bibfield  {author} {\bibinfo {author} {\bibfnamefont {J.}~\bibnamefont
  {Janszky}}\ and\ \bibinfo {author} {\bibfnamefont {A.~V.}\ \bibnamefont
  {Vinogradov}},\ }\href {\doibase 10.1103/PhysRevLett.64.2771} {\bibfield
  {journal} {\bibinfo  {journal} {Phys. Rev. Lett.}\ }\textbf {\bibinfo
  {volume} {{64}}},\ \bibinfo {pages} {2771} (\bibinfo {year}
  {1990})}\BibitemShut {NoStop}%
\bibitem [{\citenamefont {Lassig}\ and\ \citenamefont
  {Milburn}(1993)}]{LassigMilburn}%
  \BibitemOpen
  \bibfield  {author} {\bibinfo {author} {\bibfnamefont {C.~C.}\ \bibnamefont
  {Lassig}}\ and\ \bibinfo {author} {\bibfnamefont {G.~J.}\ \bibnamefont
  {Milburn}},\ }\href {\doibase 10.1103/physreva.48.1854} {\bibfield  {journal}
  {\bibinfo  {journal} {Physical Review A}\ }\textbf {\bibinfo {volume} {48}},\
  \bibinfo {pages} {1854} (\bibinfo {year} {1993})}\BibitemShut {NoStop}%
\bibitem [{\citenamefont {Chaturvedi}\ \emph {et~al.}(1998)\citenamefont
  {Chaturvedi}, \citenamefont {Milburn},\ and\ \citenamefont
  {Zhang}}]{Chaturvedi_Milburn}%
  \BibitemOpen
  \bibfield  {author} {\bibinfo {author} {\bibfnamefont {S.}~\bibnamefont
  {Chaturvedi}}, \bibinfo {author} {\bibfnamefont {G.~J.}\ \bibnamefont
  {Milburn}}, \ and\ \bibinfo {author} {\bibfnamefont {Z.}~\bibnamefont
  {Zhang}},\ }\href {\doibase 10.1103/physreva.57.1529} {\bibfield  {journal}
  {\bibinfo  {journal} {Physical Review A}\ }\textbf {\bibinfo {volume} {57}},\
  \bibinfo {pages} {1529} (\bibinfo {year} {1998})}\BibitemShut {NoStop}%
\bibitem [{\citenamefont {Radcliffe}(1971)}]{radcliffe1971}%
  \BibitemOpen
  \bibfield  {author} {\bibinfo {author} {\bibfnamefont {J.~M.}\ \bibnamefont
  {Radcliffe}},\ }\href {\doibase 10.1088/0305-4470/4/3/009} {\bibfield
  {journal} {\bibinfo  {journal} {Journal of Physics A: General Physics}\
  }\textbf {\bibinfo {volume} {4}},\ \bibinfo {pages} {313} (\bibinfo {year}
  {1971})}\BibitemShut {NoStop}%
\bibitem [{\citenamefont {Atkins}\ and\ \citenamefont
  {Dobson}(1971)}]{atkins1971angular}%
  \BibitemOpen
  \bibfield  {author} {\bibinfo {author} {\bibfnamefont {P.}~\bibnamefont
  {Atkins}}\ and\ \bibinfo {author} {\bibfnamefont {J.}~\bibnamefont
  {Dobson}},\ }\href {\doibase 10.1098/rspa.1971.0035} {\bibfield  {journal}
  {\bibinfo  {journal} {Proc. R. Soc. Lond. A}\ }\textbf {\bibinfo {volume}
  {321}},\ \bibinfo {pages} {321} (\bibinfo {year} {1971})}\BibitemShut
  {NoStop}%
\bibitem [{\citenamefont {Arecchi}\ \emph {et~al.}(1972)\citenamefont
  {Arecchi}, \citenamefont {Courtens}, \citenamefont {Gilmore},\ and\
  \citenamefont {Thomas}}]{arecchi1972atomic}%
  \BibitemOpen
  \bibfield  {author} {\bibinfo {author} {\bibfnamefont {F.}~\bibnamefont
  {Arecchi}}, \bibinfo {author} {\bibfnamefont {E.}~\bibnamefont {Courtens}},
  \bibinfo {author} {\bibfnamefont {R.}~\bibnamefont {Gilmore}}, \ and\
  \bibinfo {author} {\bibfnamefont {H.}~\bibnamefont {Thomas}},\ }\href
  {\doibase 10.1103/PhysRevA.6.2211} {\bibfield  {journal} {\bibinfo  {journal}
  {Physical Review A}\ }\textbf {\bibinfo {volume} {6}},\ \bibinfo {pages}
  {2211} (\bibinfo {year} {1972})}\BibitemShut {NoStop}%
\bibitem [{\citenamefont {Bargmann}(1947)}]{Bargmann_1947}%
  \BibitemOpen
  \bibfield  {author} {\bibinfo {author} {\bibfnamefont {V.}~\bibnamefont
  {Bargmann}},\ }\href {\doibase 10.2307/1969129} {\bibfield  {journal}
  {\bibinfo  {journal} {The Annals of Mathematics}\ }\textbf {\bibinfo {volume}
  {48}},\ \bibinfo {pages} {568} (\bibinfo {year} {1947})}\BibitemShut
  {NoStop}%
\bibitem [{\citenamefont {Novaes}(2004)}]{Novaes_2004}%
  \BibitemOpen
  \bibfield  {author} {\bibinfo {author} {\bibfnamefont {M.}~\bibnamefont
  {Novaes}},\ }\href {\doibase 10.1590/s1806-11172004000400008} {\bibfield
  {journal} {\bibinfo  {journal} {Revista Brasileira de Ensino de
  F{\'{\i}}sica}\ }\textbf {\bibinfo {volume} {26}},\ \bibinfo {pages} {351}
  (\bibinfo {year} {2004})}\BibitemShut {NoStop}%
\end{thebibliography}%
\end{document}